\documentclass[preprint,10pt]{elsarticle}
\usepackage{amsmath}
\usepackage{amssymb}
\usepackage{color}
\usepackage{mathbbol}
\usepackage{mathrsfs}
\usepackage{graphicx}
\graphicspath{{figures/}}
\usepackage{geometry}
\usepackage{overpic}
\usepackage{slashed}
\usepackage[bookmarks,bookmarksnumbered]{hyperref}
\hypersetup{colorlinks = true,linkcolor = blue,anchorcolor =red,citecolor = blue,filecolor = red,urlcolor = red,
            pdfauthor=author}
\usepackage[capitalise]{cleveref}
\usepackage{txfonts}

\journal{Computer Physics Communications}
\biboptions{numbers,sort&compress}
\usepackage{lineno}
\begin{document}
\begin{frontmatter}

\title{Rethinking the ill-posedness of the spectral function reconstruction --- \\
Why is it fundamentally hard and how Artificial Neural Networks can help}

\date{\today}

\author[a,b]{Shuzhe Shi}
\ead{shuzhe.shi@stonybrook.edu}
\author[c]{Lingxiao Wang}
\ead{lwang@fias.uni-frankfurt.de}
\author[c]{Kai Zhou}
\ead{zhou@fias.uni-frankfurt.de}

\address[a]{Department of Physics, McGill University, Montreal, Quebec H3A 2T8, Canada.}
\address[b]{Center for Nuclear Theory, Department of Physics and Astronomy, Stony Brook University, Stony Brook, New York, 11784, USA.}
\address[c]{Frankfurt Institute for Advanced Studies, Ruth Moufang Strasse 1, D-60438,Frankfurt am Main, Germany.}

\begin{abstract}
Reconstructing hadron spectral functions through Euclidean correlation functions are of the important missions in lattice QCD calculations. However, in a K\"allen--Lehmann(KL) spectral representation, the reconstruction is observed to be ill-posed in practice.
It is usually ascribed to the fewer observation points compared to the number of points in the spectral function. In this paper, by solving the eigenvalue problem of continuous KL convolution, we show analytically that the ill-posedness of the inversion is fundamental and it exists even for continuous correlation functions. We discussed how to introduce regulators to alleviate the predicament, in which include the Artificial Neural Networks(ANNs) representations recently proposed by the Authors in~[Phys. Rev. D 106 (2022) L051502]. The uniqueness of solutions using ANNs representations is manifested analytically and validated numerically. Reconstructed spectral functions using different regularization schemes are also demonstrated, together with their eigen-mode decomposition. We observe that components with large eigenvalues can be reliably reconstructed by all methods, whereas those with low eigenvalues need to be constrained by regulators.

Source code and data are publicly available at: [\href{https://github.com/ShuzheShi/SpectralFunction}{https://github.com/ShuzheShi/SpectralFunction}]
\end{abstract}


\end{frontmatter}
\newpage

\section{Introduction}
Accessing real-time dynamics of strongly interacting quantum systems requires analytic continuation of imaginary time observables in Monte Carlo methods to real time~\cite{jarrell:1996bayesian,kabanikhin:2011inverse}.
Especially in Quantum Field Theory (QFT), first-principle approaches such as lattice calculations usually are carried out in Euclidean space-time. To decode the physics of interest one has to reconstruct the spectral functions from the Euclidean correlation functions numerically computed on lattice.
In the context of high-energy nuclear physics, the proper estimation of spectral functions is essential to our understanding of properties of the hot and dense Quantum ChromoDynamics (QCD) matters which can be found in the early universe or from heavy-ion collisions. Being one important example, the QCD transport properties~\cite{Moore:2008ws, Aarts:2014nba,Itou:2020azb} can be related to the low-frequency part of spectral functions of proper current-current correlations, such as the bulk and shear viscosities needed for hydrodynamics simulations can be accessed by spectral analysis on correlation functions of the energy-momentum tensor~\cite{Moore:2008ws,Astrakhantsev:2015jta,Astrakhantsev:2017nrs}, while using the vector current for heavy quark one can correspondingly extract the heavy quark diffusion coefficient~\cite{Caron-Huot:2009ncn,Petreczky:2005nh}.
The hadronic spectrum inside the QCD medium is another example that requires spectral reconstruction and can reveal important properties of the QCD matter at finite temperature, e.g., the in-medium modification of heavy quarkonium bound states~\cite{Aarts:2013kaa,Aarts:2014cda,Burnier:2015tda} serves as ``smoking-gun'' of the quark-gluon-plasma (QGP) formation, with the peak structures' shift and broadening of the spectral function~\cite{Kim:2018yhk} reflecting the deconfinement physics of QCD at finite temperature.

In non-perturbative Monte Carlo (e.g. lattice QCD) calculations, the physical spectral functions can not be computed directly, and one would have to extract them from a finite set of correlation data~\cite{Asakawa:2000tr}. Observables of common interest include K\"allen--Lehmann(KL) correlation functions
\begin{align}
D(k) \equiv \;& 
    \int_0^\infty 
    \frac{\omega \,\rho(\omega)}{\omega^2 + k^2} 
    \frac{\mathrm{d}\omega}{\pi}, \quad k \in [0, +\infty),
    \label{eq.corr_D}
\end{align}
and the correlation function in Euclidean time. At finite temperature, $T = 1/\beta$, the mapping from the spectral function to the Euclidean correlation function given by the periodic extension of
\begin{align}
G_\beta(\tau) \equiv 
    \int_0^{\infty} 
    \frac{\cosh(\omega\tau - \omega\beta/2)}{\sinh(\omega \beta/2)}
    \rho(\omega)
    \frac{\mathrm{d}\omega}{2\pi}
    , \quad \tau \in [0, \beta].
    \label{eq.corr_G}
\end{align}
In both convolutions, while the bosonic spectral function $\rho(\omega)$ contains the physics information of interest, the correlation functions, $D(k)$ and/or $G_{\beta}(\tau)$, are the quantities that can be measured/computed directly. 
It is of high interest to extract the spectral function from the data of correlation functions. 

Extraction of the spectral function is observed to be ill-posed, however. The spectral functions are found to be highly degenerated --- one might find numerically distinctive spectral functions correspond to correlations functions that are consistent within numerical uncertainty/accuracy. The difference in the reconstructed spectral functions might lead to a difference in the interpretation of the physics meaning. As an interesting and fundamental example, it was shown that the finite temperature heavy quark interaction can be read out from the spectral analysis of the thermal Wilson loop correlation~\cite{Rothkopf:2011db,Burnier:2014ssa,Burnier:2015tda,Bala:2019cqu} or via potential model fitting to the in-medium spectroscopy~\cite{Larsen:2019bwy,Larsen:2019zqv,Larsen:2020rjk,Shi:2021qri}. Recently it is found that assumptions of different forms of spectral function can induce huge differences in the inference results for the interaction~\cite{Bala:2021fkm}.

In practical systems that $D(k)$ or $G_{\beta}(\tau)$ are only measured at finite number of $k$- or $\tau$-points and of finite precision. Eqs.~\eqref{eq.corr_D} and \eqref{eq.corr_G} are respectively discretized as 
\begin{align}
    D_i =\,& \sum_{j=1}^{N_\omega} \frac{\omega_j}{\omega_j^2 + k_i^2} \rho_j \frac{\Delta\omega_j}{\pi},
    \quad i \leq N_k,
    \label{eq.corr_Dd}
\end{align}
and
\begin{align}
    G_i =\,& \sum_{j=1}^{N_\omega} \frac{\cosh(\omega_j\tau_i - \omega_j\beta/2)}{\sinh(\omega_j \beta/2)} \rho_j \frac{\Delta\omega_j}{2\pi},
    \quad i \leq N_\tau.
    \label{eq.corr_Gd}
\end{align}
Although the problem set-up is clear and even looks simple, one usually has $N_k < N_\omega$ and/or $N_\tau < N_\omega$. In the literature, the ill-posedness is always ascribed to the non-invertible convolution matrices due to the limitation of the number of $k$- or $\tau$-points, (see e.g. ~\cite{Tripolt:2018xeo} for a useful review). 

In this paper, however, we will show analytically that the ill-posedness of the inversion of Eqs.~\eqref{eq.corr_D} and \eqref{eq.corr_G} is deeper than the limitation of observable points--- \emph{it exists even for the case that $N_k \geq N_\omega$ and/or $N_\tau \geq N_\omega$. The ill-posedness is caused by the inaccuracy of the data, no matter the measurement of spectral function is continuous or not}.
In other words, even in the relatively ideal case that one measures the correlation functions ($D$ or $G_\beta$) at any given $k\in[0,+\infty)$ or $\tau\in[0,\beta]$, but with non-zero uncertainties, no matter how small the uncertainties are, the reconstructed spectral function would bear considerable uncertainties compared to itself. Such an analytical analysis is performed by solving the eigenvalue problems. The integrals in Eqs.~\eqref{eq.corr_D} and \eqref{eq.corr_G} can be regarded as linear operators that map a continuous function, $\rho(\omega)$, to another continuous function, $D(k)$ or $G_\beta(\tau)$, and the corresponding eigenfunctions and eigenvalues can be calculated. We find that some of the eigenvalues of such operators can be arbitrarily close to zero, which makes the inversion of the integral unstable against arbitrarily small change in $D(k)$ or $G_\beta(\tau)$. The instability against the noise in correlation functions leads to the ill-posedness of the inversion.

In both continuous and discrete convolutions, the ill-posedness is due to a high degeneracy in solution space. For this reason, many regularization schemes have been proposed to break the degeneracy. 
In the classical method, the Tikhonov regularization is widely used as a Lagrange multiplier~\cite{bertero:1989linear,Tikhonov1943OnTS,tikhonov:1995numerical}. In modern lattice QCD calculations, the statistical inference with the Shannon--Jaynes entropy regulator was introduced to 
solve the problem~\cite{jarrell:1996bayesian,Asakawa:2000tr}, or named it as the maximum entropy method (MEM). In practice, it comprises prior knowledge from physical domains to regularize the inversion. Solutions of spectral functions will balance between reducing observation errors and reaching prior models~\cite{Asakawa:2000tr,Burnier:2013nla,Burnier:2014ssa}. They were validated in different scenarios with satisfying performances. Different from the above schemes, in Ref.~\cite{Wang:2021jou}, we proposed Neural Networks representations in an unsupervised automatic differentiation(AD) framework. It achieves comparable performances and its efficiency will be manifested in this article.

This paper is organized as follows. Details of the analytical calculation will be given in Sec.~\ref{sec.illposedness}. Then, methods of breaking the degeneracy will be discussed in Sec.~\ref{sec.regulator}.
Particularly, the uniqueness of the ANN construction of spectral function will be discussed in Sec.~\ref{sec.regulator.nn} and ~\ref{sec.regulator.nnp2p}.

\section{Ill-Posedness of the Continuous K\"allen--Lehmann Convolution}
\label{sec.illposedness}
\subsection{Eigenvalue Problem of Continuous Convolutions}
\label{sec.illposedness.eigenvalue}
We note that Eq.~\eqref{eq.corr_D} and \eqref{eq.corr_G} can be generalized as the Mellin convolution:
\begin{align}
	o(x) = \int_0^\infty \rho(\omega) \, K(x,\omega) \,\mathrm{d}\omega,
\end{align}
with $K(x,\omega)$ being the convolution kernel, then our aim is to retrieval $\rho(\omega)$ from finite-precise observable $o(x)$ at limited points/range of $x$. Let us focus on the K\"allen--Lehmann kernel, in which both $\rho(\omega)$ and $o(x)$ are real and of the same domain $[0,+\infty)$. Then, the Mellin convolution $\int_0^\infty [\cdot] \, K(x,\omega)\mathrm{d}\omega$ serves as a linear transformation in the Hilbert space spanned by the one-dimensional real functions with positive arguments. One can define the eigenvalue problem 
\begin{align}
\int_0^\infty \psi_s(\omega)\, K(x,\omega)\, \mathrm{d}\omega = \lambda_s \psi_s(x),
\end{align}
where $\lambda_s$ is a constant serving as the eigenvalue, and $\psi_s$ is the corresponding eigenfunction with label $s$. \textit{The invertibility of the convolution depends on the smallest value of $|\lambda_s|$ --- if $|\lambda_s|$ is zero or an arbitrarily small value, then invert-convolution is ill-posed; otherwise, if $|\lambda_s|$ has a non-vanishing lower bound, then the inversion is well-posed.} In Ref.~\cite{mcwhirter:1978numerical}, the authors found the exact solutions of the eigenvalue problem for kernels that can be expressed as a single-variable function of $x$-times-$\omega$, $K(x,\omega) = K(x\,\omega)$, which covers both Laplace and Fourier transformations. The eigenvalues were found to be
\begin{align}
\lambda^\text{[Fourier]}_{\pm,s} =\,& \pm\sqrt{\pi/2} \,, \\
\lambda^\text{[Laplace]}_{\pm,s} =\,& \pm \sqrt{\pi} \cosh^{-\frac{1}{2}}(\pi s) \,,
\end{align}
where $s \in (-\infty, +\infty)$ is the real-valued, continuous label of the eigenstate.
While $|\lambda^\text{Fourier}_{\pm,s}| = \sqrt{\pi/2}$ has a non-vanishing lower-bound,  $|\lambda^\text{Laplace}_{\pm,s}|$ can be arbitrarily small. 
Consequently, the Fourier transformation is invertible, whereas the Laplace transformation is not.

Inspired by Ref.~\cite{mcwhirter:1978numerical}, we follow similar procedures, in which eigenfunctions and eigenvalues for the KL transformation can be found by noting that
\begin{align}
\int_0^\infty x^{-1+is} \frac{1}{\pi} \frac{x\,\mathrm{d}x}{x^2+k^2} = \frac{1}{2\cosh(\pi s/2)} k^{-1+is}.
\end{align}
We can express the solutions in real functions:
\begin{align}
\psi_{+,s}(x) =\,& 
    \frac{\cos\big(s\ln (x/a)\big)}{\sqrt{\pi}\, x/a}\,,
    \label{eq.eigen_+}\\
\psi_{-,s}(x) =\,& 
    \frac{\sin\big(s\ln (x/a)\big)}{\sqrt{\pi}\, x/a}\,,
    \label{eq.eigen_-}\\
\lambda_{\pm,s} = \; & \frac{1}{2\cosh(\pi s/2)} \,.
    \label{eq.eigen_lambda}
\end{align}
where $a$ is the scaling constant of the energy unit.
Again, the label $s \in \mathbb{R}$ is continuous.
The eigenvalues of $\psi_{+,s}$ and $\psi_{-,s}$ degenerate, and we omit the $\pm$ label by denoting $\lambda_{s}\equiv \lambda_{+,s}=\lambda_{-,s}$.
The eigenfunctions~(\ref{eq.eigen_+}-\ref{eq.eigen_-}) are complete and orthonormal,
\begin{align}
&\int_{-\infty}^{+\infty} 
    \psi_{+,s}(x) \psi_{-,s}(x') \mathrm{d}s 
= 0
    \,,\\
\begin{split}
&\int_{-\infty}^{+\infty} 
    \psi_{\pm,s}(x) \psi_{\pm,s}(x') \mathrm{d}s 
=
   \frac{a}{x'} \Big[\delta\Big(\frac{x}{a}-\frac{x'}{a}\Big) \pm  \delta\Big(\frac{x}{a}-\frac{a}{x'}\Big)\Big]
    \,,
\end{split}\\
&\int_{0}^{+\infty} 
    \psi_{+,s}(x) \psi_{-,s'}(x) 
    \frac{x\, \mathrm{d}x}{a^2} 
= 0
    \,,\\
\begin{split}
&\int_{0}^{+\infty} 
    \psi_{\pm,s}(x) \psi_{\pm,s'}(x) 
    \frac{x\, \mathrm{d}x}{a^2} 
=
    \delta(s-s') \pm \delta(s+s')
    \,,
\end{split}
\end{align}
and span a complete set of basis in the Hilbert space.
Hence, for arbitrary real-valued function $f$ defined in the domain $[0,+\infty)$, one can decompose them with such a basis,
\begin{align}
f(x) =\,& \sum_{i=\pm} \int_{-\infty}^{+\infty} \frac{\mathrm{d}s}{2}
\widetilde{f}_{\pm}(s) \, \psi_{i,s}(x) \,,\label{eq.GFT_f}\\
\widetilde{f}_{\pm}(s) =\,&	\int_0^{+\infty} \frac{x\,\mathrm{d}x}{a^2} \, f(x) \, \psi_{\pm,s}(x) \,,\label{eq.GFT_b}
\end{align}
with $\widetilde{f}_{\pm}(s)$ being the coefficients of $f$ in the eigenfunction space.
Eqs.~(\ref{eq.GFT_f}--\ref{eq.GFT_b}) can be regarded as a Generalized Fourier Transformation (GFT). In the rest of this paper, we refer to $f(x)$'s as the functions in the {generalized coordinate space}, and $\widetilde{f}_{\pm}(s)$ as the function in the {generalized momentum space}.

\subsection{Formal Inversion Transformation}
\label{sec.illposedness.inversion}
Taking the GFT~(\ref{eq.GFT_f}--\ref{eq.GFT_b}) to both $\rho(\omega)$ and $D(k)$, one can rewrite the convolution equation~\eqref{eq.corr_D} as an algebra equation in the generalized momentum space,
\begin{align}
\widetilde{\rho}_\pm(s)
 = \widetilde{D}_\pm(s) \,\big/\, \lambda_{s}\,.
\label{eq.algebra_D_rho}
\end{align}
Such relation leads to a formal solution of inverting the KL convolution:
\begin{align}
\begin{split}
\rho(\omega) 
=\,& 
    \sum_{i=\pm}\int_{-\infty}^{\infty}\frac{\mathrm{d}s}{2\lambda_{s}}\psi_{i,s}(\omega)\int_0^{\infty} \frac{k\,\mathrm{d}k}{a^2} \psi_{i,s}(k) D(k)
\\=\,&
    \int_0^{\infty} \frac{k\,\mathrm{d}k}{a^2}  D(k)
    \int_{-\infty}^{\infty}\frac{\mathrm{d}s}{2\lambda_{s}}
    \sum_{i=\pm} \psi_{i,s}(\omega) \psi_{i,s}(k)
\\=\,&
    \int_0^{\infty} \frac{D(k) \, \mathrm{d}k}{\pi \, \omega}
    \int_{-\infty}^{\infty} \mathrm{d}s\,
    \cos(s\ln\frac{k}{\omega})\cosh(\frac{\pi s}{2})\,.
\end{split}
\label{eq.inversion_equation}
\end{align}
The integral 
\begin{align}
    I(x) \equiv \int_{-\infty}^{\infty} \mathrm{d}s\,
    \cos(s x)\cosh(\frac{\pi s}{2})\,,
    \label{eq.inverse_kl_integral}
\end{align}
serves as the analytic form of the inversion kernel in the Backus--Gilbert method~\cite{BackusGilbert}. However, it does not converge in the region that $\lambda_{s}\to0$.
In the ideal case where we know the exact form of $D(k)$, we can exploit the properties based on regularization outlined in~\ref{sec.reg_BackusGilbert} --- given arbitrary function $f(x)$ defined on the real axis, $x \in \mathbb{R}$, its convolution with $I$ results in
\begin{align}
\begin{split}
&   \int_{-\infty}^{\infty}
    f(x)
    I(x-x_0) \mathrm{d}x
=
    2\pi\,\sum_{n=0}^{\infty}
    \frac{(-\pi^2/4)^n}{(2n)!}f^{(2n)}(x_0)\,.
\end{split}
\label{eq.delta_drv}
\end{align}
Further more, if we know its analytic continuation in the complex plane, $x\in\mathbb{C}$, we can further simplify the precedent equation as
\begin{align}
\begin{split}
&   \int_{-\infty}^{\infty}
    f(x)
    I(x-x_0) \mathrm{d}x
=
    \pi\,\Big(f(x_0+\frac{\pi i}{2})
    + f(x_0-\frac{\pi i}{2}) \Big)\,,
\end{split}
\end{align}
and recover the well-known optical theorem~\cite{zee:2010quantum},
\begin{align}
&   \rho(\omega) 
=
    -2\,\text{Im}[D(i\omega)]\,.
\label{eq.optical}
\end{align}
and the small-frequency behavior of the spectral function, which is related to the transport conductivity~\cite{Ding:2015ona,Ratti:2018ksb},
\begin{align}
\sigma \propto&
    \lim_{\omega \to 0} \frac{\rho(\omega)}{\omega}
=
    -2 D'(0)\,.
\end{align}
Detailed proof of these two relations can be found in~\ref{sec.reg_BackusGilbert}.

\subsection{Null-Modes in the Inversion}\label{sec.illposedness.null_mode}
\begin{figure}[!hbtp]
    \centering
    \includegraphics[width=0.32\textwidth]{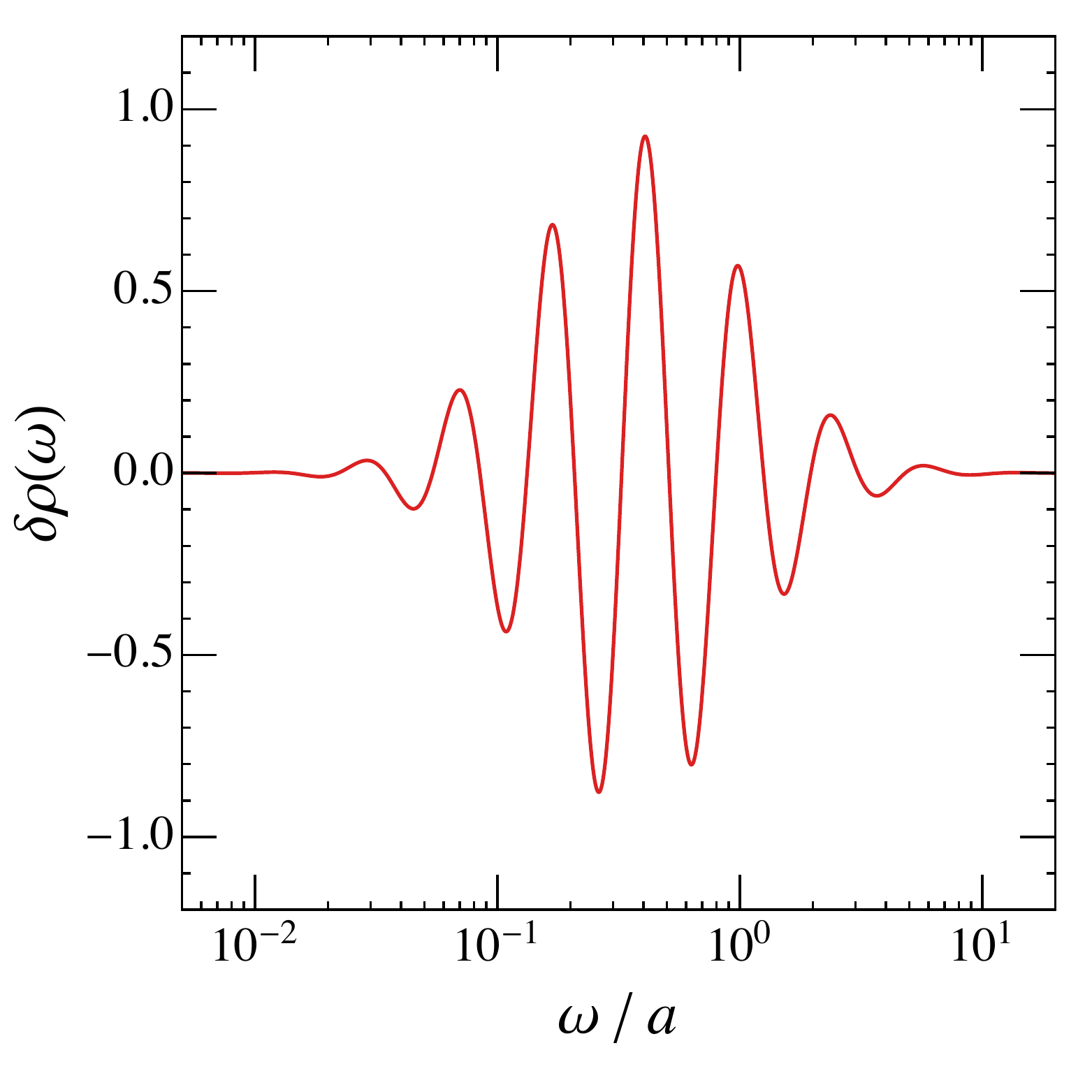}
    \includegraphics[width=0.32\textwidth]{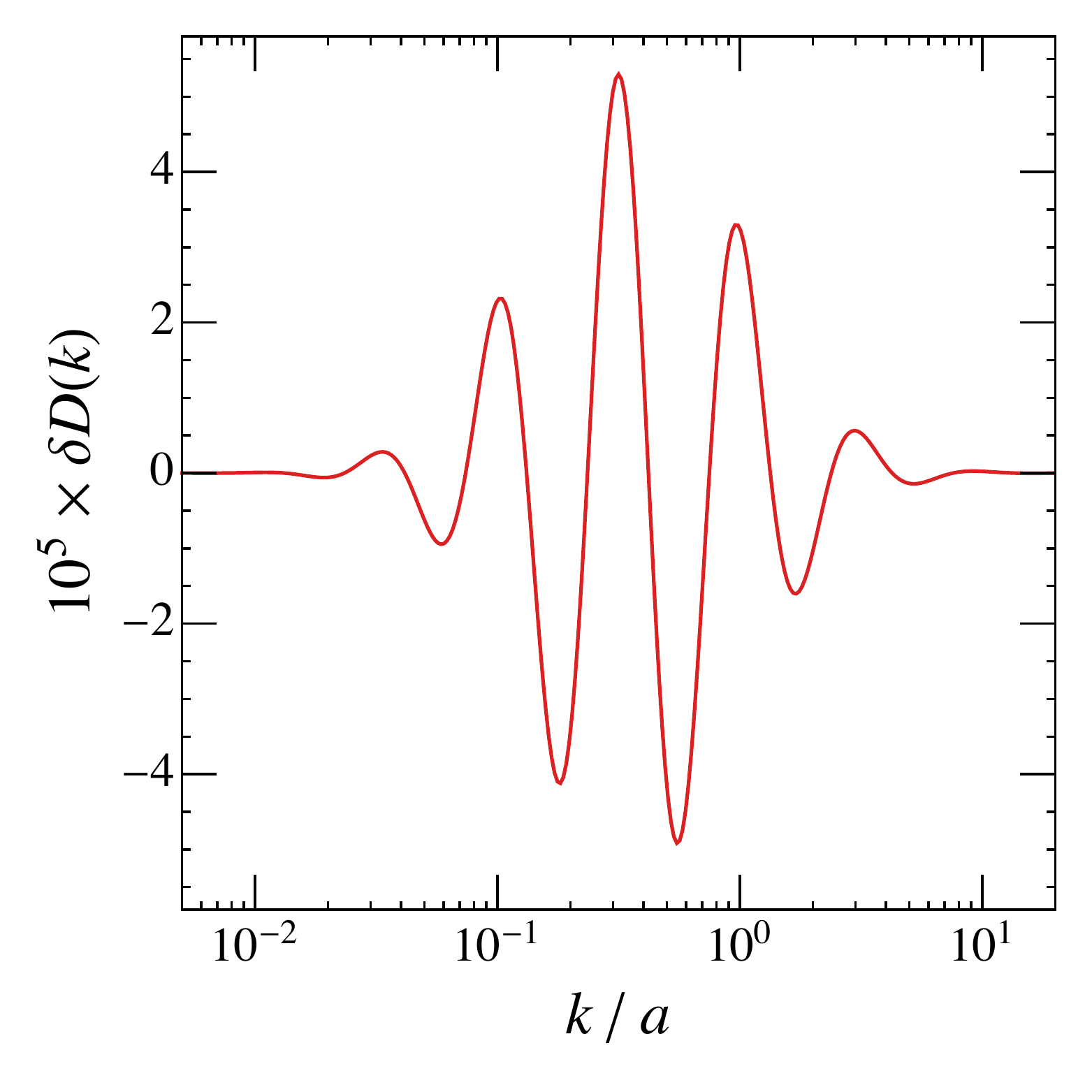}
    \includegraphics[width=0.32\textwidth]{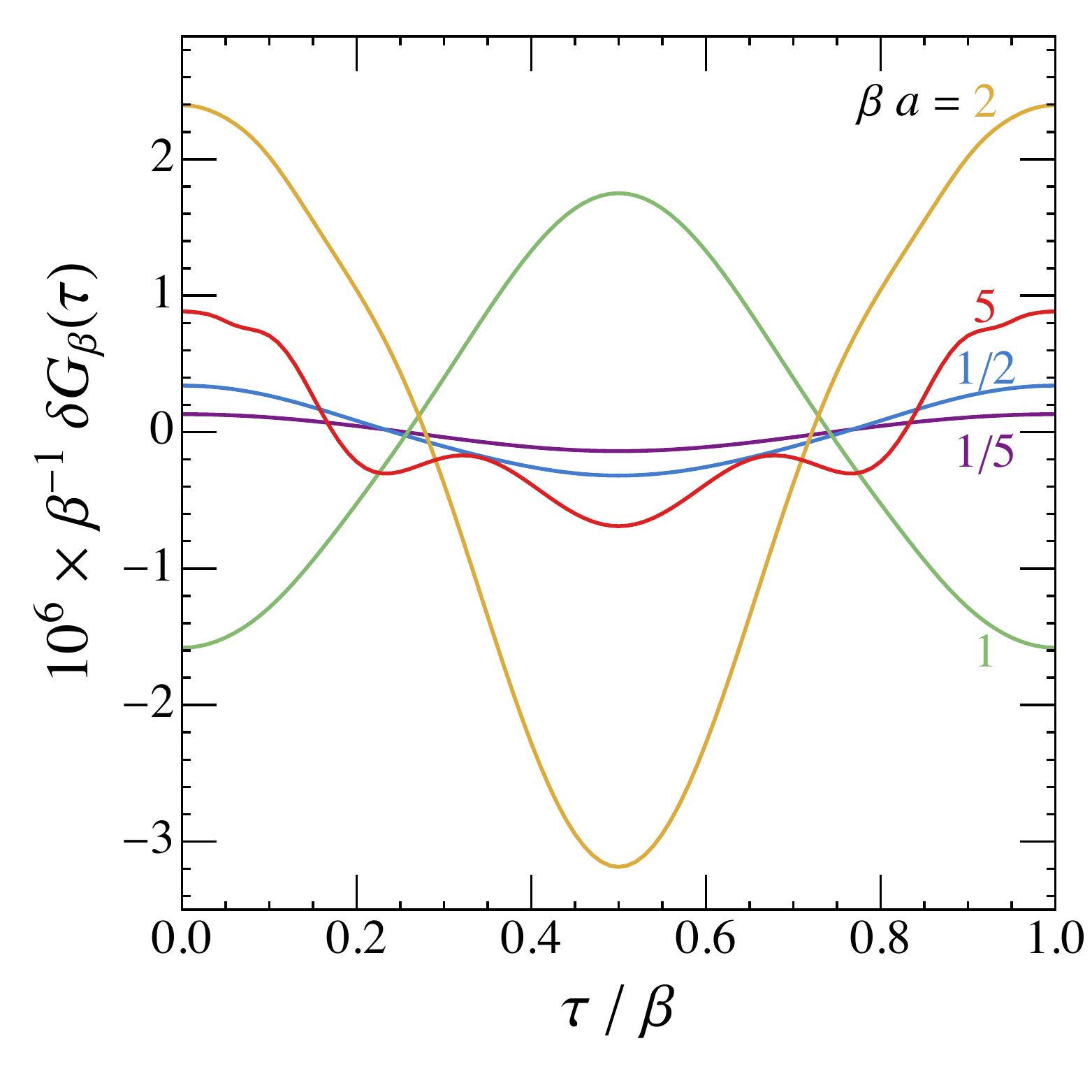}
\caption{Example of null mode (left) taking \protect{Eq.~\eqref{eq.null_gaussian}} with $s_0 = 7$ and $\sigma=1$. Corresponding perturbation in correlation functions, $\delta D(\tau)$ and $\delta G_\beta(\tau)$, is shown in the middle and right panels, respectively. In the right panel, purple, blue, green, yellow, and red curves correspond to $\beta= 1/5a$, $1/2a$, $1/a$, $2/a$, and $5/a$, respectively.}
    \label{fig.null}
\end{figure}

In the precedent subsection, we have shown that how we can exploit the GFT to invert the KL convolution in the ideal case where we know the exact form as well as the analytic continuation of $D(k)$. In this subsection, we will further demonstrate how the inversion on the KL convolution becomes ill-posed in the practical situations where the correlation functions are numerically calculated/measured with finite precision. The issue can be found from the algebra equation~\eqref{eq.algebra_D_rho} between coefficients of $\rho$ and $D$ in the eigen-space.
A small perturbation in the correlation function, $\delta \widetilde{D}_{\pm}(s) = \Delta(s)$, could lead to a huge change in the spectral function, $\delta \widetilde{\rho}_{\pm}(s) =  2 \cosh(\pi s/2) \Delta(s)$, at the large-$s$ region.
In other words, the large-$s$ components of $\rho$ can hardly be constrained by the measurement of $D(\tau)$, and one always needs a \textit{prior} knowledge of that. Hereof, we refer to the large-$s$ components that can not be constrained by the correlation data as \emph{null-modes}.

Null-modes of the KL convolution kernel are also null-modes for the correlation function in Euclidean time~\eqref{eq.corr_G}. We note that
\begin{align}
\frac{\cosh(\omega\tau - \omega\beta/2)}{\sinh(\omega \beta/2)}
= \sum_{n=-\infty}^{\infty} e^{\frac{2\pi n \tau }{\beta}i} \frac{2 \omega}{\omega^2 + n^2 \frac{4\pi^2}{\beta^2}},
\end{align}
hence,
\begin{align}
    G_\beta(\tau) =
    \sum_{n=-\infty}^{\infty} e^{\frac{2\pi n \tau }{\beta}i} D(2\pi n/ \beta)\,.
    \label{eq.Gtau_vs_Dk}
\end{align}
The null-modes in $\rho(\omega)$ we discussed above, which lead to negligible change in $D(k)$ for all $k$, would consequently lead to negligible change in $G_\beta(\tau)$.

To get some intuition of the null-modes, we take the Gaussian function concentrated at large-$s$ region as an example,
\begin{align}
\begin{split}
&\frac{1}{\sqrt{2\pi}\sigma} \int_{-\infty}^{\infty} 
    \psi_{\pm,s} e^{-\frac{(s-s_0)^2}{2\sigma^2}}\mathrm{d}s
=
     e^{-\frac{\sigma^2}{2}\ln^2\frac{x}{a}} \psi_{\pm,s_0}(x)\,,
\end{split}\label{eq.null_gaussian}\\
     ~\nonumber\\
\begin{split}
& \frac{1}{\sqrt{2\pi}\sigma} \int_{-\infty}^{\infty} 
    \frac{\psi_{\pm,s}(x)}{2\cosh(\pi s/2)} e^{-\frac{(s-s_0)^2}{2\sigma^2}} \mathrm{d}s 
\\\approx \;&
\frac{1}{\sqrt{2\pi}\sigma} 
\int_{-\infty}^{\infty} 
    \frac{\psi_{\pm,s}(x)}{\exp(\pi s/2)} e^{-\frac{(s-s_0)^2}{2\sigma^2}} \mathrm{d}s 
\\= \;&
    e^{\frac{\pi^2\sigma^2}{8} - \frac{\pi s_0}{2}} 
    e^{-\frac{\sigma^2}{2}\ln^2\frac{x}{a}} \psi_{\pm, s_0-\pi \sigma^2/2 }(x)\,.
\end{split}
\end{align}

It means that by introducing $\delta \widetilde{\rho}_i(s) = \text{Gaussian}(s_0,\sigma; s)$ with large enough $s_0$, one can alter $\rho(\omega)$ significantly [$\sim\mathcal{O}(1)$] but retain $D(\tau)$ to be within $\mathcal{O}(e^{-\pi s_0/2})$ accuracy. We take the perturbation $\delta\rho(\omega)$ as in Eq.~\eqref{eq.null_gaussian} with taking the $+$ sign in the subscript and $s_0 = 7$ and $\sigma=1$, and evaluate the corresponding correlation functions, $D(k)$ and $G_\beta(\tau)$. We present the results in Fig.~\ref{fig.null}. We find an $\mathcal{O}(1)$ perturbation in $\rho(\omega)$ would lead to $\mathcal{O}(10^{-5})$ difference in $D(k)$ and $G_\beta(\tau)$. Therefore, one cannot constrain such a perturbation when the correlation functions are of finite accuracy large than $10^{-5}$. The magnitude of $\delta D(k)$ and $\delta G_\beta(\tau)$ will be further suppressed when $s_0$ increases in Eq.~\eqref{eq.null_gaussian}. As will be presented later in Sec.~\ref{sec.regulator.practical} with practical examples, the differences between numerically reconstructed spectral functions and the corresponding truth value are observed to be the difference of the functions in large generalized momentum space.

It shall be worth noting that similar null mode behaviors can be observed for discrete systems~\eqref{eq.corr_Dd} and \eqref{eq.corr_Gd} with large enough $N_\omega$ and $N_k$ or $N_\tau$. For discrete systems, one usually performs the singular value decomposition method to numerically solve the eigenvalues and eigenvectors of the convolution matrices. The eigenvectors corresponding to the nearly-vanishing eigenvalues correspond to the null modes discussed here. The analytical results given in the present work provide insights for how the null modes behave. This might be valuable in inspiring future designs of methods to regularize the ill-posed problem.

\section{Constraining the Degeneracy by Regulators}\label{sec.regulator}
\begin{figure}[htbp!]
    \centering
    \includegraphics[width= 0.8 \textwidth]{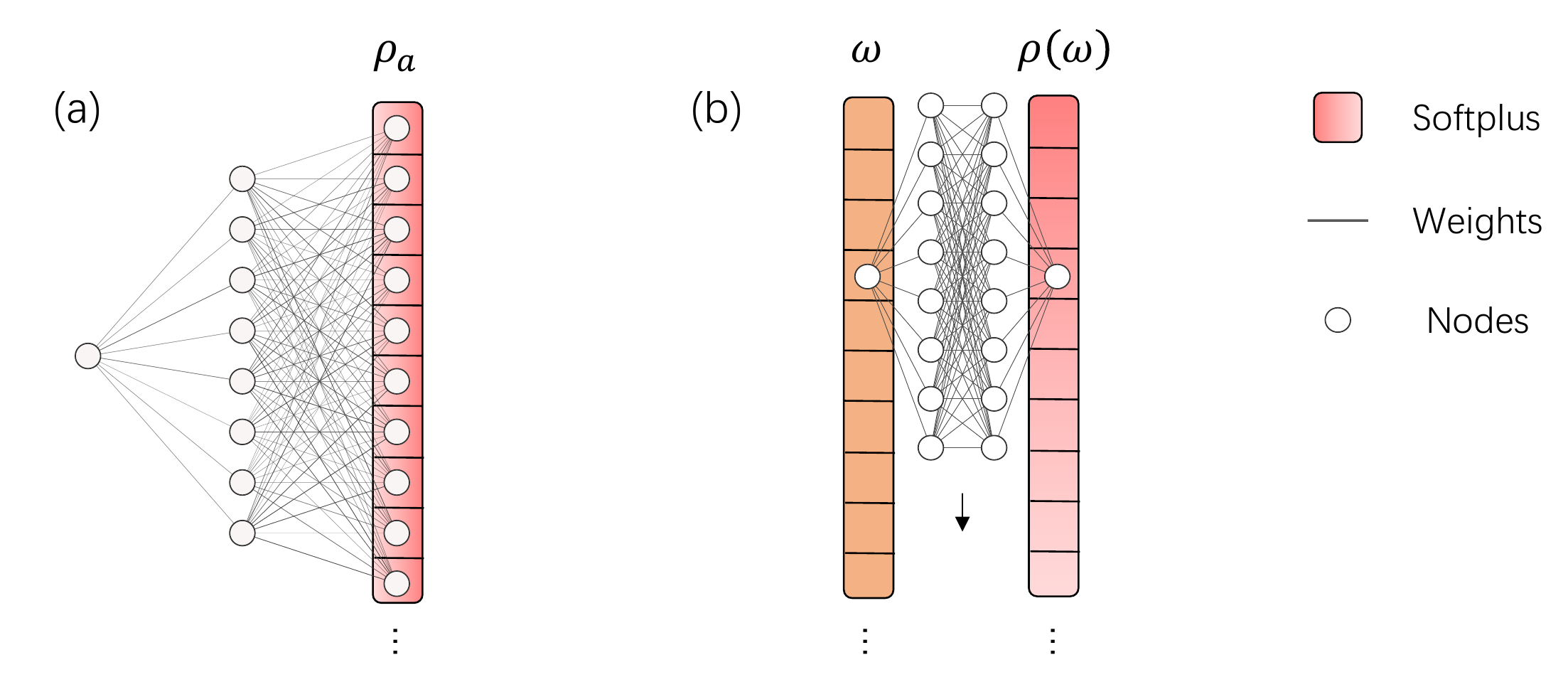}
    \caption{ANNs representations of spectral functions. (a) \texttt{NN}. Neural networks have outputs as a list representation of the spectral function values $\rho_a$. (b) \texttt{NN-P2P}. Neural networks have input and output nodes as $(\omega,\rho(\omega))$ pairwise.}
    \label{fig.arch}
\end{figure}

As preceding Sections show analytically, reconstructing the spectral function from real-valued, noisy KL correlation functions is fundamentally ill-posed. To break the degeneracy of the solution to the inverse problem, one would have to introduce regulators on the spectral function. In existing literatures, the regulator is usually formulated as entropy terms, which are functionals of the spectral function. Recently in Ref.~\cite{Wang:2021jou}, we proposed two methods beyond traditional entropy terms, in which Artificial Neural Networks (ANNs) representations are used to regularize the solution. Their representations are shown in Fig.~\ref{fig.arch}. This novel approach employs ANNs to represent the spectral function or its value at a given list of $\omega$ points. For the sake of convenience, the former scheme is annotated as \texttt{NN-P2P} whereas the latter is named as \texttt{NN}. We also note that the former --- i.e., parameterizing a continuous function with a deep neural network --- is a common practice in the machine learning community, and also widely adopted in the high energy physics community, e.g., in the parton distribution function reconstruction~\cite{Forte:2002fg,Karpie:2019eiq}. Exploiting the smoothness property of neural networks, the reconstructed spectral functions are found to be smooth and reasonably agree with the ground truth.

In this Section, we first present a pedagogical review of some commonly employed methods to regularize the null-modes, including 
Truncated Singular Value Decomposition(TSVD) method~\cite{doi:10.1137_0911028,Hansen_1992,CHEN2017297}, Tikhonov(TK) regulator~\cite{Tikhonov1943OnTS,tikhonov:1995numerical}, the Shannon–Jaynes entropy deployed in the Maximum Entropy Method(MEM)~\cite{jarrell:1996bayesian,annurev.aa.24.090186.001015}, and the Bayesian Reconstruction(BR) method~\cite{Burnier:2013nla}. 
Then in Sec.~\ref{sec.regulator.nn} and \ref{sec.regulator.nnp2p}, we show that solution of spectral function using ANN representation is unique. 

Without loss of generality, we focus on discrete convolution kernel, i.e., the transformation between spectral function and correlation function can be represented as 
\begin{align}
    D_i = \sum_{a=0}^{N_\omega} K_{ia} \rho_a \Delta\omega,
    \quad 1 \leq i \leq N\,,
\end{align}
with the convolution matrix $K_{ia} \equiv K(x_i,\omega_a)$.
Hereof, we use $a,b,c,d,\cdots$ to label the index of spectral function and/or network parameters, and $\cdots i,j,k,l \cdots$ for correlation function. Also, to avoid confusion, all summations signs will be spelled-out explicitly.

Quite often, the convolution matrix is a rectangular matrix that can not be diagonalized through orthogonal transformation.
However, one can always perform the singular value decomposition (SVD) to the convolution matrix,
\begin{align}
    K_{ia} = \sum_{b,j} U_{ab}\Sigma_{bj}V_{ij}
\end{align}
with $U$ and $V$ being orthogonal,
\begin{align}
&   \sum_a U_{ab} U_{ab'} = \delta_{bb'}\,,\quad
    \sum_b U_{ab} U_{a'b} = \delta_{aa'}\,,\\
&   \sum_i V_{ij} V_{ij'} = \delta_{jj'}\,,\quad
    \sum_j V_{ij} V_{i'j} = \delta_{ii'}\,,
\end{align}
whereas $\Sigma$ is semi-diagonal,
\begin{align}
\Sigma_{bj} = 
    \left\{ \begin{array}{ll}
    \sigma_j\delta_{bj}\,, & b\leq N_\text{rank},  \\
    0\,,     & b > N_\text{rank}.
    \end{array} \right.
\end{align}
$N_\text{rank} \leq N$ is the rank of $K$.
One can easily see that
\begin{align}
\begin{split}
    \sum_{a}K_{ia}U_{ab} 
=\,& 
    \sum_{j} \Sigma_{bj}V_{ij}\,,
\end{split}
\end{align}
hence,
\begin{align}
    \sum_{a}K_{ia}U_{ab} = 0, \quad
    \forall \, b > N_\text{rank}.
    \label{eq.null}
\end{align}

For arbitrary vector $\{f_a\}$, one can decompose
\begin{align}
    f_a = \sum_{b=1}^{N_\omega} c_b U_{ab},
\end{align}
where the coefficient is given by
\begin{align}
    c_b = \sum_{a=1}^{N_\omega} f_a U_{ab}\,.
\end{align}

With SVD, one can find 
\begin{align}
\begin{split}
D_i 
=&
    \sum_{b,j} c_{b}\Sigma_{bj}V_{ij} \Delta\omega\,,
\end{split}
\end{align}
hence
\begin{align}
\sum_{i=1}^{N} D_i V_{ik}
=\,&
   \sigma_{k} c_{k} \Delta\omega\,,
    \label{eq.tsvd_0}
\end{align}
which determines the first $N_\text{rank}$ coefficients. $c_b$ with $N_\text{rank} < b \leq N$ are not determined due to vanishing $\sigma_b$, whereas $N<b$ are not constrained by Eq.~\eqref{eq.tsvd_0}. The Truncated Singular Value Decomposition (TSVD) method~\cite{doi:10.1137_0911028,Hansen_1992,CHEN2017297} was proposed to assume that all the unconstrained coefficients vanish, and 
\begin{align}
\rho_a = 
    \sum_{b=1}^{N_\text{rank}} c_b U_{ab}
=
    \sum_{b=1}^{N_\text{rank}} \sum_{i=1}^{N} \frac{D_i^\text{obs} V_{ib} U_{ab}}{\sigma_b \Delta\omega}\,.
    \label{eq.tsvd}
\end{align}
Although Eq.~\eqref{eq.tsvd} gives a solution, out-of-the degenerated ones, there is no guarantee that it is the physically correct one. Indeed, TSVD is disfavored in practice since it usually results in fast-oscillating spectral functions.

To eliminate the null-modes and ensure smoothness, one routinely includes extra regulators to ``punish'' the null-modes and break the degeneracy (see e.g.~\cite{kaipio2006statistical} and the references therein). Generally speaking, to solve the ambiguity caused by the null-modes, one needs to minimize not only the distance between observations and reconstructions but also the regularization terms. Modern regularization techniques can be formatted in a statistical inference manner to be maximizing the Bayesian Posterior,
\begin{align}
    P(\rho|D,I) = \frac{P(D|\rho,I) P(\rho|I)}{P(D|I)},
\end{align}
where the likelihood reads
\begin{align}
    P(D|\rho,I) = e^{-\chi^2/2}\,,
\end{align}
and the prior 
\begin{align}
    P(\rho|I) = e^{\mathcal{S}[\rho]}
\end{align}
carries our prior knowledge of the spectral function.
Maximization of the Bayesian Posterior is equivalent to minimization of the loss function in below as in the traditional methods
\begin{align}
    J \equiv \frac{\chi^2}{2}
    - \mathcal{S}[\rho]\,.
    \label{eq.regularization}
\end{align}
We refer the readers to Ref.~\cite{Bertero:1988eh} for a review of regularization methods in general, and 
e.g. Refs.~\cite{KRYZHNIY2004618,Brianzi_1991} for regularized Laplace Transformation. Attempts of regularized inversion of the KL kernel are also discussed in~\cite{Dudal:2013yva}.

With these preparations in linear-algebra, we are ready to discuss the uniqueness of regularized reconstruction, 
which maximizes the Posterior
\begin{align}
    P = e^{-\frac{\chi^2}{2} + \mathcal{S}} \equiv e^{-J}\,.
\end{align}
The $\chi^2$ function is the uncertainty-weighted difference between the data ($D_i^\mathrm{obs}$) and the correlation function computed from the reconstructed spectral function, at corresponding momentum point ($D(k_i) \equiv D[\rho(\omega)](k_i)$),
\begin{align}\label{eq.chisquare}
    \chi^2 \equiv \sum_{i,j=1}^{N} C_{ij}^{-1}
    (D_i^\mathrm{obs} - D_i) (D_j^\mathrm{obs} - D_j)\,,
\end{align}
where $C^{-1}$ is the inverse covariance matrix.
For later convenience, we denote that
\begin{align}
\Delta_i \equiv 
    - \frac{\delta \chi^2/2}{\delta D(k_i)}
    = \sum_j C_{ij}^{-1}
    (D_j^\mathrm{obs} - D(k_j)),
\end{align}
which is a functional of $\rho(\omega)$ through $D(k_i)$.

\subsection{Tikhonov, the Shannon–Jaynes Entropy, and Bayesian Reconstruction Regularization terms}
For pedagogical reason, let us first review the commonly employed entropy-based regulation methods.
They include Tikhonov(TK) regulator~\cite{Tikhonov1943OnTS,tikhonov:1995numerical},  the Shannon–Jaynes entropy deployed in the Maximum Entropy Method(MEM)~\cite{jarrell:1996bayesian,annurev.aa.24.090186.001015}, and the Bayesian Reconstruction(BR) method~\cite{Burnier:2013nla}, all of which is designed to minimize the difference of the spectral function ($\rho$) to a defaulted model ($\text{DM}$),
\begin{align}
\mathcal{S}_\text{TK} =\,& 
    -\frac{\alpha}{2}\sum_{a=1}^{N_\omega} (\rho_a - \text{DM}_a)^2 \Delta\omega \,,\\
\mathcal{S}_\text{MEM} =\,& 
    \alpha \sum_{a=1}^{N_\omega} \Big(\rho_a - \text{DM}_a - \rho_a\ln\frac{\rho_a}{\text{DM}_a}\Big) \Delta\omega\,,\\
\mathcal{S}_\text{BR} =\,& 
    \alpha \sum_{a=1}^{N_\omega} \Big(1- \frac{\rho_a}{\text{DM}_a} + \ln\frac{\rho_a}{\text{DM}_a}\Big) \Delta\omega\,.
\end{align}
In traditional TK reconstruction, $\alpha$ is a hyper parameter needs to be carefully chosen~\cite{kaipio2006statistical}, whereas this parameter is integrated out in MEM and BR, and the reconstructed spectral function is $\rho(\omega) = \int P(\alpha|D,\text{DM}) \rho_\alpha(\omega) \mathrm{d}\alpha$. $\rho_\alpha$ is the reconstructed spectral function given $\alpha$, and $P(\alpha|D,\text{DM}) = \int P(\alpha,\rho|D,\text{DM}) \mathcal{D}\rho_\alpha$ which is a functional integral, $P(\alpha,\rho|D,\text{DM})\propto 
e^J P(\alpha)$ is the joint possibility.
Although the $\alpha$-dependence will be removed through integration over $P(\alpha)$, calculations of $P(\alpha|D,\text{DM})$ need careful handling of the $\alpha$. For MEM, one usually takes $P(\alpha) = 1/\alpha$, which has the simplest scale-invariant form following the Jeffreys prior~\cite{jeffreys1946invariant}. More details can also be found in Ref.~\cite{jarrell:1996bayesian,annurev.aa.24.090186.001015,Pavarini2012CorrelatedEF}. Different from the scale-invariant consideration, in BR approach~\cite{Burnier:2013nla}, a fully $\alpha$-independent approximation was introduced as a constant, i.e., $P(\alpha) = 1$.

The inclusion of the entropy term -- for any given $\alpha$ -- leads to a unique solution of the spectral function. Noting that the functional variation $\frac{\delta J}{\delta \rho(\omega)} = 0$ vanishes when $J$ is minimized and the Posterior is maximized, we find the optimal spacial functions respectively satisfy
\begin{align}
\begin{split}
   \rho^\text{TK}_{a} - \text{DM}_a
=\,& 
   \frac{1}{\alpha} \sum_{i}
   \Delta_i^\text{TK} K_{ia}\,,
\end{split}\\
\begin{split}
    \ln \frac{\rho^\text{MEM}_{a}}{\text{DM}_a}
=\,& 
   \frac{1}{\alpha} \sum_{i} 
   \Delta_i^\text{MEM} K_{ia}\,,
\end{split}\\
\begin{split}
   \frac{1}{\text{DM}_a} - \frac{1}{\rho^\text{BR}_a}
=\,& 
   \frac{1}{\alpha} \sum_{i}
    \Delta_i^\text{BR} K_{ia}\,.
\end{split}
\end{align}

One can expand the difference between  the spectral function and the default model in a complete set of $N_\omega$-dimensional vectors,
\begin{align}
    \rho^\text{TK}_{a} =\,& \text{DM}_a + \sum_{b=1}^{N_\omega} c_b^\text{TK} U_{ab}\,,\\
    \rho^\text{MEM}_{a} =\,& \text{DM}_a \exp\bigg(\sum_{b=1}^{N_\omega} c_b^\text{MEM} U_{ab}\bigg)\,,\\
    \rho^\text{BR}_{a} =\,& \bigg[(\text{DM}_a)^{-1} -  \bigg(\sum_{b=1}^{N_\omega} c_b^\text{BR} U_{ab}\bigg)^{-1}\bigg]^{-1}\,,
\end{align}
and the unitarity of $U$-matrix yields that
\begin{align}
\begin{split}
c_b^\text{TK} =\,&
   \sum_{a=1}^{N_\omega} (\rho^\text{TK}_{a} -  \text{DM}_a) U_{ab}
=
   \frac{1}{\alpha} \sum_{i}
   \Delta_i^\text{TK} \sum_{a=1}^{N_\omega} K_{ia} U_{ab} \,,
\end{split}\label{eq.coef_tk}\\
\begin{split}
c_b^\text{MEM} =\,&
   \sum_{a=1}^{N_\omega} \ln\frac{\rho^\text{MEM}_{a}}{  \text{DM}_a} U_{ab}
=
   \frac{1}{\alpha} \sum_{i}
   \Delta_i^\text{MEM} \sum_{a=1}^{N_\omega}K_{ia} U_{ab} \,,
\end{split}\label{eq.coef_mem}\\
\begin{split}
c_b^\text{BR} =\,&
   \sum_{a=1}^{N_\omega} \Big(\frac{1}{\text{DM}_a} - \frac{1}{\rho^\text{BR}_a}\Big)U_{ab}
=
   \frac{1}{\alpha} \sum_{i}
   \Delta_i^\text{BR} \sum_{a=1}^{N_\omega}K_{ia} U_{ab}  \,.
\end{split}\label{eq.coef_br}
\end{align}
These equations are the self-consistent equations that determine the coefficients uniquely. In practical MEM calculations, the equation is usually solved using Bryan's sequential least square quadratic programing(SLSQP) method~\cite{bryan1990maximum} which reconstructs MEM spectral function based on coefficients with $b \leq  N_\text{rank}$. We note that there is a controversy of whether the solution obtained by the Bryan's method is complete or not. Refs.~\cite{Rothkopf:2011ef,Rothkopf:2019ipj,Rothkopf:2020qqt} argued that the singular subspace is incomplete, whereas Ref.~\cite{Asakawa:2020hjs} claims the opposite.
Indeed, if one combines Eq.~\eqref{eq.null} with Eqs.~(\ref{eq.coef_tk} -- \ref{eq.coef_br}), the solution is automatically constrained in the singular subspace,
\begin{align}
c_b^\text{X} =\,&
   \frac{1}{\alpha} \sum_{l} \Sigma_{bl} \sum_{i}
   \Delta_i^\text{X} V_{il}  \,,
\end{align}
hence,
\begin{align}
    c_b^\text{X} = 0,\quad\forall\, b > N_\text{rank}.
\end{align}
With the above analysis, we conclude that Bryan's TSVD method is complete and accurate, assuming infinite numerical precision.
In practice, on the other hand, we notice that some open-source numerical programs based on Bryan's optimization method truncate out the small but non-vanishing eigenvalues in $K_{ia}$\footnote{Such eigen-modes correspond to the null-modes discussed in Sec.~\ref{sec.illposedness}}, which makes $N_\text{rank} \ll N$. It is important to keep those modes in the updating process in order to reach the true maximum of the Posterior. Nevertheless, we note that including all eigen-modes does not necessarily induce a smooth reconstruction of spectral functions. See Figs.~\ref{fig.performance1} and~\ref{fig.performance2} for a concrete example in which MEM result with complete bases becomes unstable against the noise in correlation functions. A more detailed analysis of the instability is given in~\ref{sec.mem_unstable}. Therefore, in unstable cases, truncation might be needed and can be regarded as an extra regularization of the spectral function, and the criteria of the truncation shall be spelled out explicitly.

On a separate note, one can formally summarize TK, MEM, and BR regularization schemes into a uniform one, which introduces an auxiliary function, $f(\omega_a)$, to re-parametrize the spectral function,
\begin{align}
\rho_a =\,& R^{(1,0)}(f_a,\omega_a),\\
\mathcal{S} =\,& 
    - \alpha \sum_a\big(f_a \rho_a -  R(f_a,\omega_a)\big) \Delta\omega \label{eq.reg},
\end{align}
where $R(f,\omega)$ is a function that embeds the physics prior knowledge. $R^{(1,0)}(f,\omega) \equiv \partial R(f,\omega) / \partial f$ is the first-order derivative of $R$ with respect to its first argument, and serve as a functional translating $f(\omega)$ to $\rho(\omega)$.
The minimum solution of the loss function $J = \chi^2/2 - \mathcal{S}$ yields,
\begin{align}
0 = \frac{\delta  J}{ \delta f_a}
    &= \alpha f_a \frac{\delta\rho_a}{\delta f_a} \Delta\omega - \sum_i \Delta_{i} K_{i,a}\frac{\delta\rho_a}{\delta f_a} \Delta\omega,
\end{align}
hence
\begin{equation}
    f_a = \frac{1}{\alpha} \sum_i \Delta_{i} K_{i,a}.
\end{equation}
Therefore, the auxiliary function $f_a$ belongs to the singular space, $f_a = \sum_{b=1}^{N_\mathrm{rank}} c_b U_{ab}$, where $c_b \equiv \alpha^{-1} \sum_{i,j} \Delta_i V_{ij} \Sigma_{bj}$. TK, MEM, and BR regulators respectively correspond to
\begin{align}
R_\text{TK}(f,\omega) =\,& 
    \frac{1}{2}f^2 - \text{DM}(\omega) f\,,\\
R_\text{MEM}(f,\omega) =\,& 
    \text{DM}(\omega)\, (e^f-1)\,,\\
R_\text{BR}(f,\omega) =\,& 
    -\ln\big(1-\text{DM}(\omega) f\big)\,.
\end{align}

\subsection{Neural Network Construction of the Spectral Function List (\texttt{NN})}\label{sec.regulator.nn}

In Ref.~\cite{Wang:2021jou}, we propose a neural network construction of the $\{\rho_a\equiv\rho(\omega_a)\}$ vector, called \texttt{NN}-architecture, which generates the list as
\begin{align}
\rho_a =\,& 
    \text{DM}_a \, \sigma^{(l)}(f^{(l)}_a)\,,\label{eq.rho_nn}\\
\begin{split}
f_a^{(n)} =\;& \sigma^{(n)}(x_a^{(n)})\,,\\
x_a^{(n)} =\,& \sum_{b} W^{(n)}_{ab} f^{(n-1)}_b,
\end{split} \quad 
\begin{split}
    &a=1,2,\cdots,N^{(n)},\\
    &n=1,2,\cdots,l,
\end{split}
\label{eq.x_iteration}
\end{align}
where $\text{DM}_a$ is the defaulted model, $l$ is referred to as the \emph{number of layers}, $1\leq n\leq l-1$ are the labels of \emph{hidden layers}, whereas $n=0$ the \emph{input layer}, $n=l$ the \emph{output layer}. Index  $a$($b$) labels the $a$-th($b$-th) \emph{neuron} at the $n$-th($n-1$-th) layer, and $N^{(n)}$($N^{(n-1)}$) is called the \emph{width} of such layer. $W^{(n)}_{a,b}$ are the \emph{weights}\footnote{We have restricted that all biases are zero to ensure uniqueness of the solution.}, whereas $x^{(n)}_a$ and $f^{(n)}_a$ the input and output of a neuron, $\sigma^{(n)}$ the non-linear activation function. By construction, $N^{(0)}=1$, $N^{(l)}=N_\omega$, and the input layer $a_1^{(0)}=1$.

To obtain analytical results, we show the proof of uniqueness for the simplified set-up using linear activation functions for the hidden layers, $f^{(n)}_a = x^{(n)}_a$ for $n\le l$, and allow arbitrary activation function for the output layer $\sigma^{(l)}(x) = \sigma(x)$. We further denote that $f_a \equiv f^{(l)}_a$.
For later convenience, we define 
\begin{align}
G_a \equiv\;&
    \frac{1}{2\alpha\Delta\omega} \frac{\partial \chi^2}{\partial f_a}
    = \alpha^{-1}\sum_{i} \Delta_i K_{ia} \text{DM}_a \sigma'(f_a),
\end{align}
which are functionals of $f_a$. Natural set-up of neuron networks also contain a $L_2$ loss term which regulates the magnitude of the weights,
\begin{align}
    L_2 \equiv \alpha \Delta\omega \sum_{l,a,b} \big(W^{(l)}_{ab}\big)^2.
    \label{eq.l2}
\end{align}

Derivative of the loss-function with respect to the last-layer weights reads
\begin{align}
 &   \frac{\partial}{\partial W_{ab}^{(l)}}\frac{\chi^2 + L_2}{2\Delta\omega}
=
    \alpha W_{ab}^{(l)} - x^{(l-1)}_{b} \alpha G_a\,,
\end{align}
and its vanishing yields the solution
\begin{align}
    W_{ab}^{(l)} = x^{(l-1)}_{b} G_a\,.
    \label{eq.W_iteration_l}
\end{align}
Similarly, derivative with respect to the weights of other layers gives
\begin{align}
W_{bc}^{(l-1)} =\,&
     x^{(l-2)}_{c}\sum_{a} G_a W^{(l)}_{ab}\,, \\
W_{cd}^{(l-2)} =\,&
     x^{(l-3)}_{d}\sum_{a,b} G_a W^{(l)}_{ab} W_{bc}^{(l-1)}\,,\\
     \vdots\;&\nonumber\\
W_{fg}^{(n)} =\,&
     x^{(n-1)}_{g}\sum_{a,b,c,\cdots,e} G_a W^{(l)}_{ab} W_{bc}^{(l-1)} \cdots W^{(n+1)}_{ef}\,,
\\\vdots\;&\nonumber\\
W_{h1}^{(1)} =\,&
     \sum_{a,b,c,\cdots,g} G_a W^{(l)}_{ab} W_{bc}^{(l-1)} \cdots W^{(2)}_{gh}\,.\label{eq.W_iteration_1}
\end{align}
To solve this equation set, we introduce an auxiliary quantity 
\begin{align}
    A_f^{(n)} \equiv \sum_{a,b,c,\cdots,e} G_a W^{(l)}_{ab} W_{bc}^{(l-1)} \cdots W^{(n+1)}_{ef}\,,
\end{align}
with $A^{(l)}_f = G_f$. The iteration equations (\ref{eq.W_iteration_l}--\ref{eq.W_iteration_1}) can be summarized as
\begin{align}
    W^{(n)}_{fg} =\,& A_f^{(n)} x^{(n-1)}_{g}\,.\label{eq.W_iteration}
\end{align}
We further define the norms as 
\begin{align}
    \lVert x^{(n)}\rVert \equiv\;& \Big(\sum_a \big(x^{(n)}_a\big)^2\Big)^{1/2}\,,\\
    \lVert A^{(n)}\rVert \equiv\;& \Big(\sum_a \big(A^{(n)}_a\big)^2\Big)^{1/2}\,,
\end{align}
Plugging Eq.~\eqref{eq.W_iteration} into Eq.~\eqref{eq.x_iteration} and the iteration property of $A^{(n)}_f$, we find
\begin{align}
x^{(n)}_f =\,& 
    \sum_g W_{fg}^{(n)} x^{(n-1)}_g
=
    A_f^{(n)} \lVert x^{(n-1)}\rVert^2\,\label{eq.iteration_xA}\\
A^{(n)}_f =\,& 
    \sum_e A_e^{(n+1)} W_{ef}^{(n+1)}
=   x_f^{(n)} \lVert A^{(n+1)}\rVert^2
    \,,\label{eq.iteration_Ax}
\end{align}
It is not hard to find
\begin{align}
1 =\,&
    \lVert A^{(n+1)}\rVert \; \lVert x^{(n-1)}\rVert \,,\\
\lVert x^{(n)}\rVert =\,&
    \lVert A^{(n)}\rVert \; \lVert x^{(n-1)}\rVert^2 \,.
\end{align}
Based on these two equations, we can show that the norms of $x^{(n)}$'s form a geometric sequence
\begin{align}
    \frac{\lVert x^{(n+1)}\rVert}{\lVert x^{(n)}\rVert} = \frac{\lVert x^{(n)}\rVert}{\lVert x^{(n-1)}\rVert} = {const.}
\end{align}
Noting that $\lVert x^{(0)}\rVert = 1$, we find $\lVert x^{(n)}\rVert = \lVert x^{(l)}\rVert^{{n}/{l}} = \lVert f \rVert^{{n}/{l}}$. Taking $n=l$, Eq.~
\eqref{eq.iteration_xA} yields that
\begin{align}
    f_a = G_a \lVert x^{(l-1)}\rVert^2 = G_a \lVert f \rVert^{\frac{2l-2}{l}}\,.
\end{align}
Expressing the terms explicitly, we find the self-consistent equation for the unique solution of $f_a$ (hence $\rho_a$),
\begin{align}
   \frac{f_{a}/\sigma'(f_a)}{\big(\sum_b f_b^2\big)^{\frac{l-1}{l}}} = \frac{\text{DM}_a}{\alpha}\sum_{i} \Delta_i K_{ia} \,.
   \label{eq.nn_solution}
\end{align}
When using \texttt{SoftPlus} activation function for the output layer, $\sigma_\text{softplus}(f) = \ln(1+e^{f})$, the solution satisfies
\begin{align}
   \frac{(1+e^{-f_a}) f_{a} }{\big(\sum_b f_b^2\big)^{\frac{l-1}{l}}} = \frac{\text{DM}_a}{\alpha}\sum_{i} \Delta_i K_{ia} \,.
   \label{eq.nn_solution_sp}
\end{align}
Compared to TK, MEM, and BR methods, the self-consistent equation obtained here shares some similarity that $ \frac{f_{a}}{\text{DM}_a \sigma'(f_a)}\big(\sum_b f_b^2\big)^{\frac{1-l}{l}}$ remains in the singular space and hence is unique. On the other hand, it contains a non-local term $\big(\sum_b f_b^2\big)^{\frac{l-1}{l}}$ when $l>1$ and heavily entangles points at different $\omega_a$'s.

For \texttt{NN} without hidden layer ($l=1$), $f_a = W^{(1)}_{a1}$, 
the $L_2$-regularization corresponds to an effective entropy term,
\begin{align}
\begin{split}
\mathcal{S}_{\text{L2}} \equiv\;&
    -\frac{L_2}{2}
    = -\frac{\alpha \Delta\omega}{2}\sum_a f_a^2
=
    -\frac{\alpha \Delta\omega}{2}\sum_a \big(\ln(e^{\frac{\rho_a}{\text{DM}_a}}-1)\big)^2\,,
\end{split}
\end{align}
and Eq.~\eqref{eq.nn_solution} can be re-expressed as,
\begin{align}
\frac{f_a}{\text{DM}_a \sigma'(f_a)} 
=   \frac{1}{\alpha}\sum_{i} \Delta_i K_{ia}\,.
    \label{eq.nn_solution_0}
\end{align}

\begin{figure}[!htbp]\centering
\includegraphics[width=0.45\textwidth]{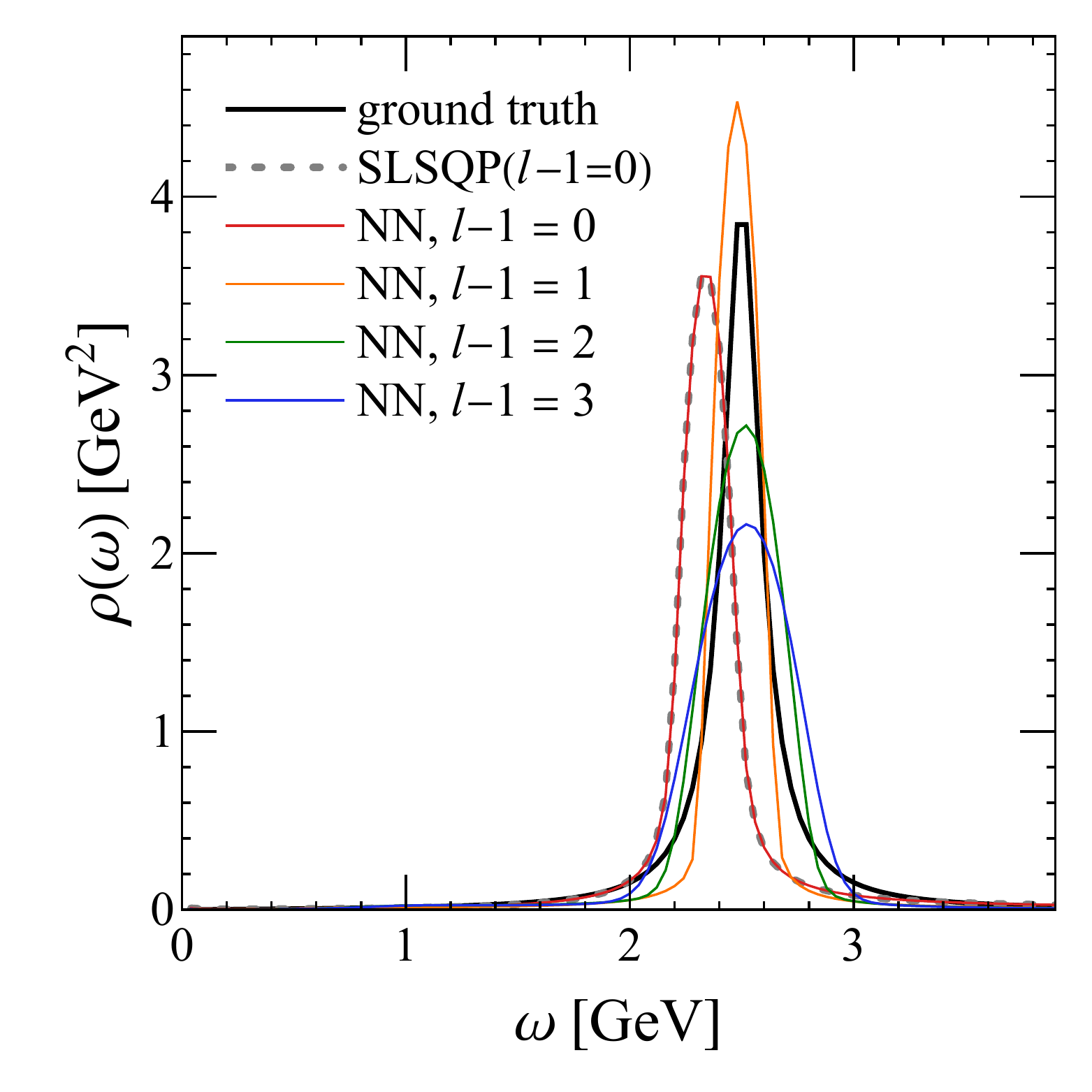}\quad
\includegraphics[width=0.45\textwidth]{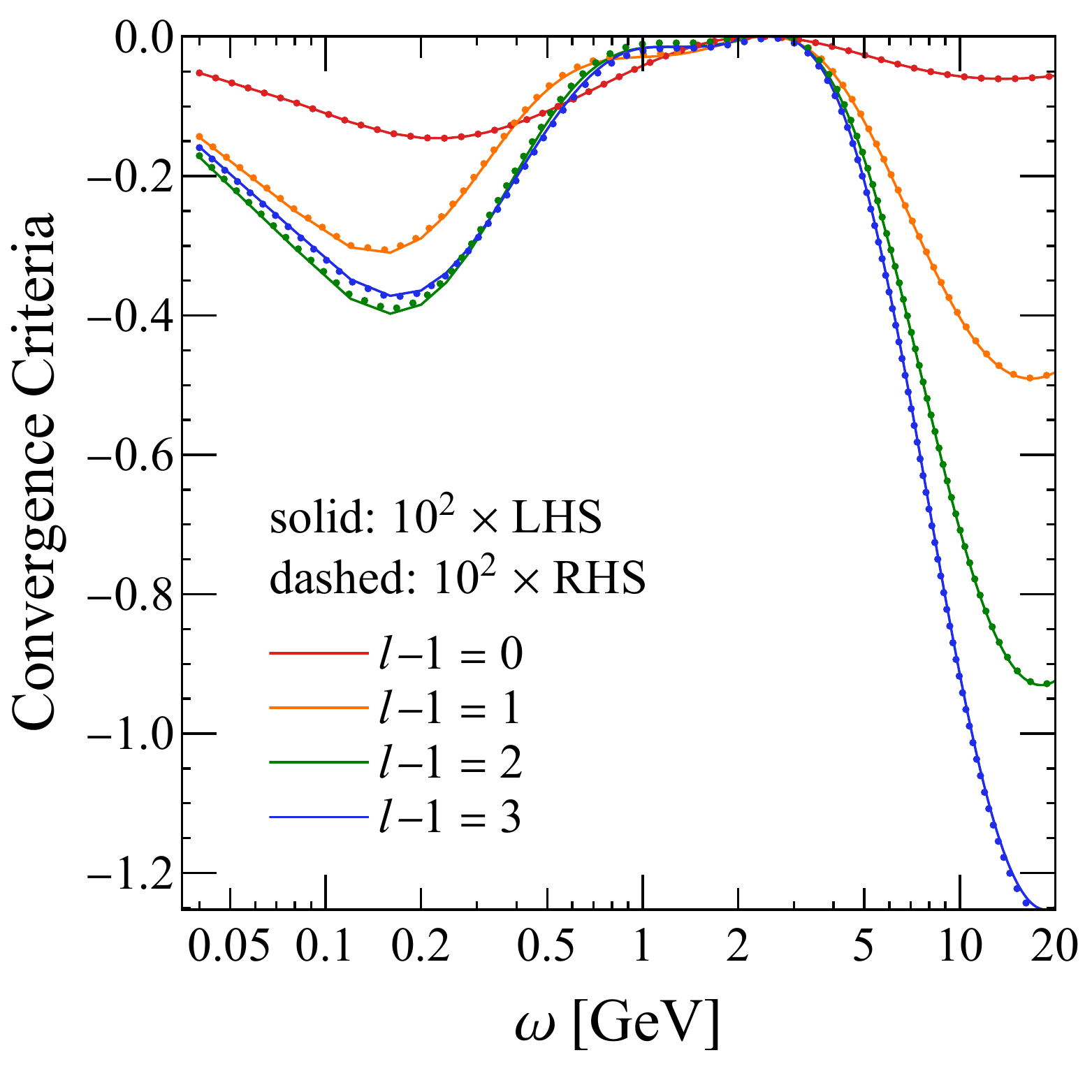}
\caption{(Left) Comparison of spectral function obtained from \texttt{NN}-architecture without hidden layer (red) and corresponding solution using Bryan's sequential least square quadratic programing (SLSQP) method (gray dashed) under the regulator~\protect{\eqref{eq.l2}}. Ground truth of the spectral function is represented by the black curve whereas \texttt{NN} results with $l-1=1$, $2$, $3$ are represented by orange, green, and blue curves, respectively. (Right) Left-hand-side (solid) and right-hand-side (dashed) of the convergence criteria~\protect{\eqref{eq.convergence_criteria}} for the \texttt{NN} result towards the corresponding unique solution.
Red, orange, green and blue curves respectively represent \texttt{NN}-architecture with $l-1=0$, $1$, $2$, and $3$ hidden layers.
\label{fig.nn_test}}
\end{figure}

Let us check that we do obtain the desired unique solution given by the self-consistent equations when using the aforementioned \texttt{NN}-architecture.
We start from a known spectral function with a Breit--Wigner peak
$\rho(\omega) = \frac{4 A \Gamma\omega}{\left(M^{2}+\Gamma^{2}-\omega^{2}\right)^{2}+4 \Gamma^{2} \omega^{2}}$,
with $A = 1$, $\Gamma = 0.1$~GeV, $M = 2.5$~GeV, and compute the corresponding KL correlation functions $D(k_i)$ at $k_i = i\times\Delta k$, with $i=1,2,\cdots,100$, and  $\Delta k=0.2$~GeV. Then, we reconstruct the spectral function $\rho(\omega)$ for points $\omega_a = a\times \Delta\omega$, with $a=1, 2, \cdots, 500$, and $\Delta\omega=0.04$~GeV. The default model is set to be $\text{DM}_a=1~\text{GeV}^2$. The inverse covariance matrix in the $\chi^2$ function~\eqref{eq.chisquare} is chosen to be $C^{-1}_{ij}=\delta_{ij}$, with $\alpha = 10^{-6}$ for the regulator parameter in Eq.~\eqref{eq.l2}.
In Fig.~\ref{fig.nn_test}~(left) we adopt \texttt{NN}-architecture without hidden layer and compare the obtained spectral function to the corresponding solution of the self-consistent equation~\eqref{eq.nn_solution_0} by using Bryan's sequential least square quadratic programming (SLSQP) method~\cite{bryan1990maximum}. We observe good agreement between these two approaches. 
For \texttt{NN} with non-zero hidden layers, their unique solutions~\eqref{eq.nn_solution_sp} are not able to be obtained using the Bryan's SLSQP method. 
Therefore, on the right panel we compare the left-hand-side and right-hand-side of the following convergence criteria
\begin{align}
\alpha\,f_{a}\,(1+e^{-f_a}) \equiv \;& 
    \text{DM}_a \Big(\sum_b f_b^2\Big)^{\frac{l-1}{l}} \sum_{i} \Delta_i K_{ia}\,,
    \label{eq.convergence_criteria}
\end{align}
using \texttt{NN}-architecture with $l-1=0$(red), $1$(orange), $2$(green), and $3$(blue) hidden layers after $\sim10^5$ training steps. We observe good agreement between the left-hand-side and right-hand-side of the convergence criteria.
Both of these examinations indicate that we do reach the desired unique solution.

We note that above self-consistent solutions obtained in this subsection are based on \texttt{linear} activation function in the hidden layers. Putting other activation functions, the $L_2$ regularization puts different constraints on $x^{(l)}_b$, and allows a different structure of self-consistent equations that might not be able to explicitly represent. Therefore, the general type of \texttt{NN} set-up provides more flexibility -- compared to the traditional $\mathcal{S}[\rho]$ regulators -- in regularizing the spectral function. Finally, we note that the \texttt{NN}-architecture can easily be amended to a common list representation of the spectral function --- one can adopt a set-up without any hidden layer, relieve the $L_2$ regulator of the weights, and introduce additional loss terms into the loss function.

\subsection{DNN Representation of the Spectral Function (\texttt{NN-P2P})}\label{sec.regulator.nnp2p}
Deep Neural Networks (DNNs) can generally provide an unbiased, yet flexible enough, parameterization to approximate arbitrary function/functional relations. A mathematically strict proof is provided by the universal approximation theorem~\cite{Leshno93multilayerfeedforward, Kratsios_2021}.
A DNN essentially acts as a piece-wise representation of a function/functional, and regularizations of its parameters ensure the smoothness of the function, which grants the advantage of avoiding over-fitting~\cite{2017arXiv170610239W,rosca:2020case}.

Recalling that the key to constraining the null-modes is to ensure the smoothness of the function and eliminate the oscillating modes, we can exploit the fact that DNN automatically imposes the smoothness condition. Hence, in Ref.~\cite{Wang:2021jou} we proposed to represent the spectral function by a DNN, which defines the spectral function in the manner of iterative non-linear function composition,
\begin{align}
\rho(\omega) =\,& 
    \text{DM}(\omega) f^{(l)}_1(\omega),\\
\begin{split}
f^{(n)}_a(\omega) =\,&
    \sigma^{(n)}(x^{(n)}_a(\omega)),\\
x^{(n)}_a(\omega) =\,&
    b^{(n)}_a + \sum_{b=1}^{N^{(n-1)}} W^{(n)}_{a,b} f^{(n-1)}_b(\omega),
\end{split}
\quad
\begin{split}
    &a=1,2,\cdots,N^{(n)},\\
    &n=1,2,\cdots,l,
\end{split}
\end{align}
where $N^{(0)}=N^{(l)}=1$ and $a_1^{(0)}(\omega)=\omega$ by definition. Compared to the \texttt{NN} set-up~(\ref{eq.rho_nn}, \ref{eq.x_iteration}), the major difference is that \texttt{NN} represents the spectral function as a discrete list at given $\omega$'s, whereas \texttt{NN-P2P} represents the function as a continuous function of $\omega$. Hence, all $x^{(n)}_a(\omega)$'s and $f^{(n)}_a(\omega)$'s are functions of $\omega$. Besides, the \emph{biases} $b^{(n)}_a$ can be non-vanishing.
We reconstruct the spectral function by optimizing the network parameters in order to minimize the loss function, defined as the summation of the chi-square function and the $L_2$-regularization of the network weights,
\begin{align}
    J_\text{NN-P2P} = \chi^2 + \frac{\alpha}{2} \sum_{n,a,b} \big(W^{(n)}_{a,b}\big)^2\,.
\end{align}
This method is referred to as \texttt{NN-P2P} in~\cite{Wang:2021jou}, which means a Neural-Network construction of a Point-to(2)-Point function.
We note that general steps of solving regularized ill-posed inverse problems using DNN are outlined in~\cite{Adler_2017}.

It shall be worth noting that there is a different class of works employing Neural Networks to solve the inverse problem~\cite{Kades:2019wtd,2018PhRvB..98x5101Y,2020PhRvL.124e6401F,2020InvPr..36f5005L,Chen:2021giw}. In these works, the authors start from a set of known spectral functions and compute the corresponding correlation functions, and use this data set to train the Neural Networks approximating the functional mapping from correlation functions to spectral functions. In such practices, regulators are implicitly introduced in the preparation of the data set. However, one shall be cautious for the risk that priors for the specific form of spectral functions might have been introduced in the preparation of the training data set.

To show the uniqueness of \texttt{NN-P2P} representation, we take a wide-enough network with one hidden layer, which adopts the \texttt{relu} activation function [$\sigma_\text{relu}(x)=(x+|x|)/2$] for all layers, followed by an extra \texttt{softplus} activation of the output, as well as the defaulted model as $\text{DM}(\omega) = \omega$.
Taking such a set-up, we express the spectral function as
\begin{align}
\rho(\omega) =\,& 
    \omega \ln(1+e^{f(\omega)})\,,\\
f(\omega) =\,& 
    \sum_{a=1}^{N_W}{W_a W'_a \sigma_\text{relu}(\omega - \omega_a)}\,,
    \label{eq.f_dnn}
\end{align}
where $W_a$ and $W'_a$ are the weights, and the biases are absorbed by $\omega_a$. It is not hard to see that Eq.~\eqref{eq.f_dnn} provides a piece-wise linear interpolation joining the points discretized in $\{\omega_a\}$, whereas $W_a W'_a$ is the change of slope for the two segments connected to $\omega_a$. Hence, the introduction of $L_2$-regularization,
\begin{align}
    L_2 = \alpha \Delta\omega \sum_{a} (W_a^2 + W_a'^2),
\end{align}
is essentially constraining the second-order derivative and therefore constrains the oscillations.
The parameter-gradient of the loss functions read
\begin{align}
   \frac{\partial}{\partial \omega_a}\frac{\chi^2 + L_2}{2\Delta\omega}
=\,&
    W_a W'_a \sum_{i}\Delta_i I_{i,a}\,,\\
   \frac{\partial}{\partial W_a}\frac{\chi^2 + L_2}{2\Delta\omega}
=\,&
    \alpha W_a - W'_a \sum_{i}\Delta_i (J_{i,a}-\omega_a I_{i,a})\,,\\
   \frac{\partial}{\partial W'_a}\frac{\chi^2 + L_2}{2\Delta\omega}
=\,&
    \alpha W'_a - W_a \sum_{i}\Delta_i (J_{i,a}-\omega_a I_{i,a})\,,
\end{align}
where 
\begin{align}
\begin{split}
I_{i,a} \equiv\;&
    \int_{\omega_a}^{\omega_\text{max}} K(k_i,\omega) (1-e^{-\frac{\rho(\omega)}{\omega}}) \omega\mathrm{d}\omega\,,
\end{split}\\
\begin{split}
J_{i,a} \equiv\;&
    \int_{\omega_a}^{\omega_\text{max}} K(k_i,\omega)(1-e^{-\frac{\rho(\omega)}{\omega}})\omega^2 \mathrm{d}\omega\,.
\end{split}
\end{align}
The vanishing of derivatives indicates that
\begin{align}
0 =\,& \sum_{i}\Delta_i I_{i,a}\,,\\
\alpha =\,& \Big|\sum_{i}\Delta_i J_{i,a}\Big|\,,
\end{align}
together with $|W_a| = |W'_a|$. While these conditions are different from the typical uniqueness condition given by e.g., Tikhonov regulator form, one can see that there are $3\times N_W$-independent conditions --- all $3\times N_W$ parameters are constrained and there is no unfixed degree of freedom. Besides, we note that the uniqueness holds even for vanishing $\alpha$. Oscillations of the represented function have been avoided when limiting the number of intermediate neurons~\cite{2017arXiv170610239W,rosca:2020case}.

\subsection{Practical Performance of Different Methods}\label{sec.regulator.practical}
While all aforementioned methods are effective in breaking the degeneracy and damping the oscillating null-modes, they are not guaranteed to provide the correct inversion function, at the presence of finite noise. In this subsection, we show the practical performance of different methods in reconstructing the spectral functions. 

\begin{figure*}[!h]\centering
\includegraphics[width=0.32\textwidth]{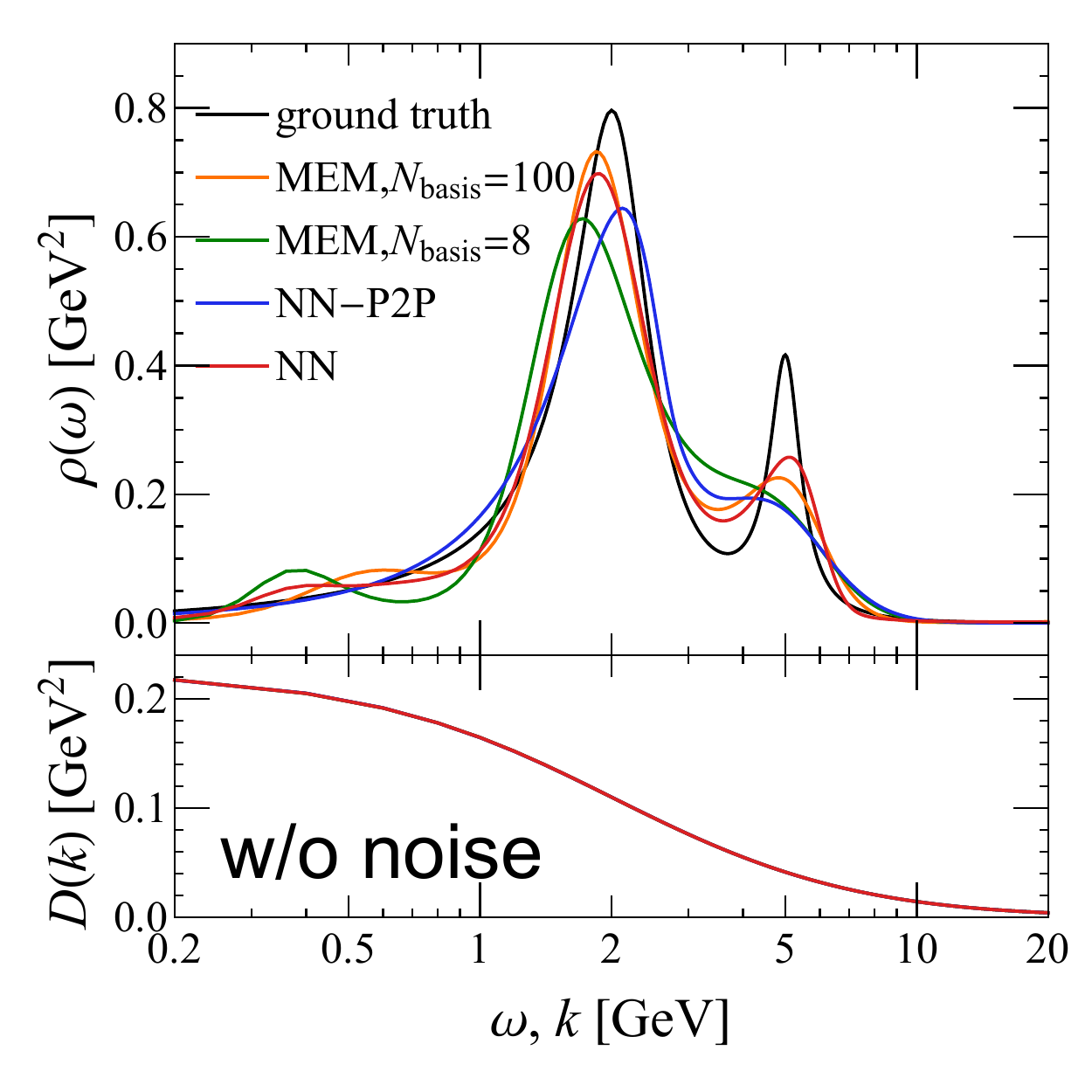}
\includegraphics[width=0.32\textwidth]{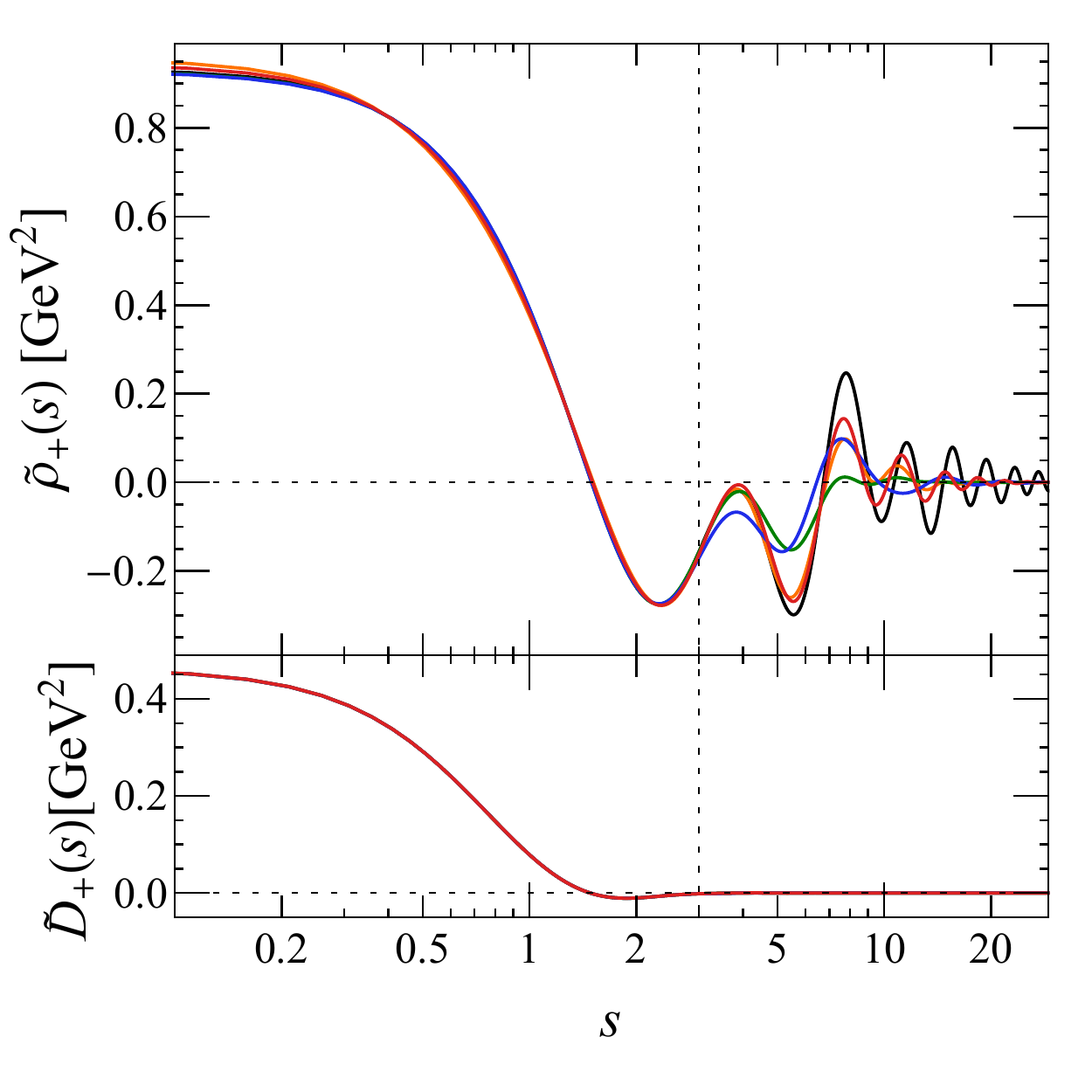}
\includegraphics[width=0.32\textwidth]{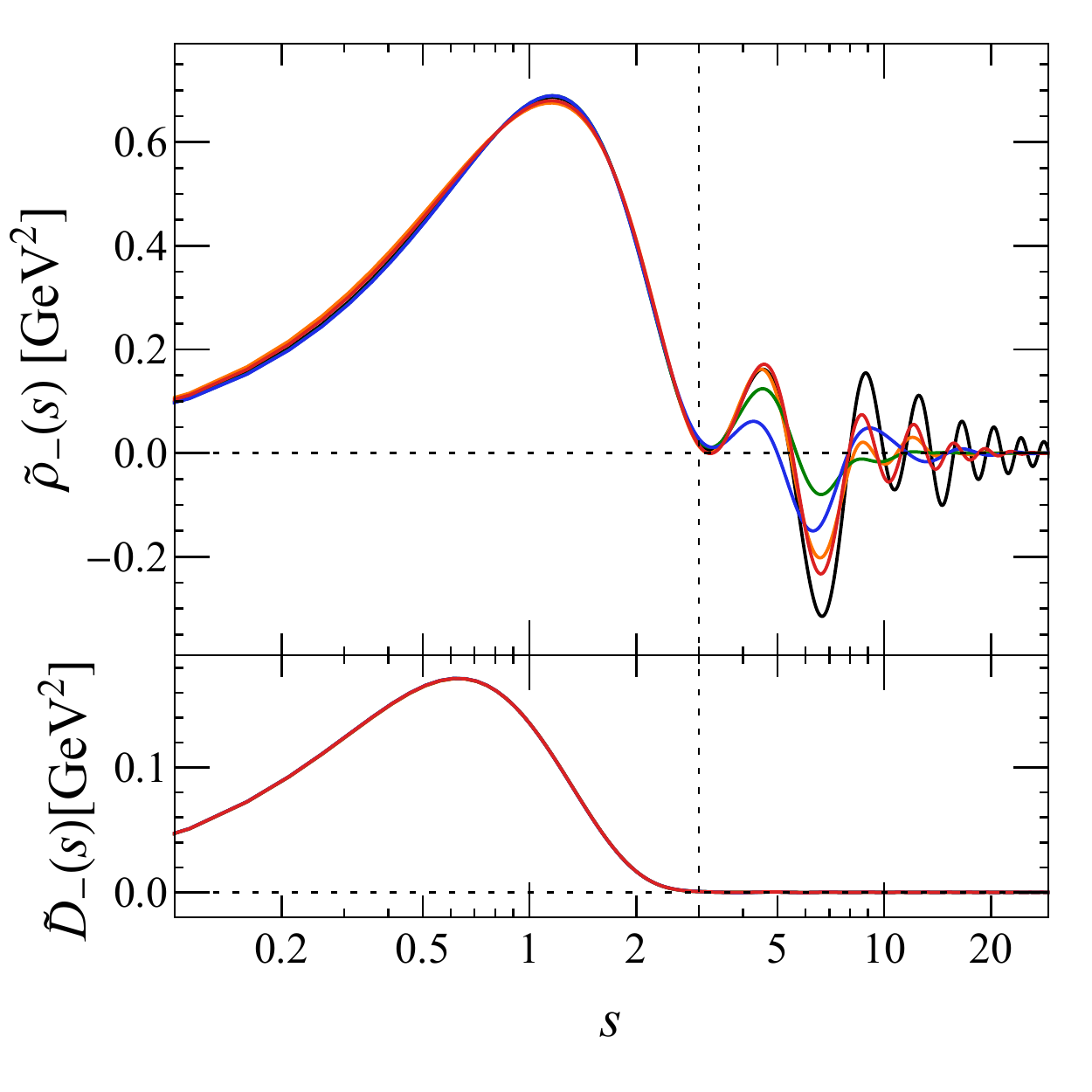}
\caption{Spectral functions using different reconstruction methods (upper panels) and their corresponding KL correlation functions (lower panels) in the generalized coordinate space (left) and generalized momentum space (middle and right). 
Black curves are for the ground truth using Breit--Wigner spectral~\protect{\eqref{eq.breit_wigner}}. Numerically reconstructed functions using \texttt{NN}, \texttt{NN-P2P}, and MEM using $N_\text{basis}=8$ and $N_\text{basis}=100$ basis are represented by red, blue, green and orange curves, respectively.\label{fig.performance1}}
\includegraphics[width=0.32\textwidth]{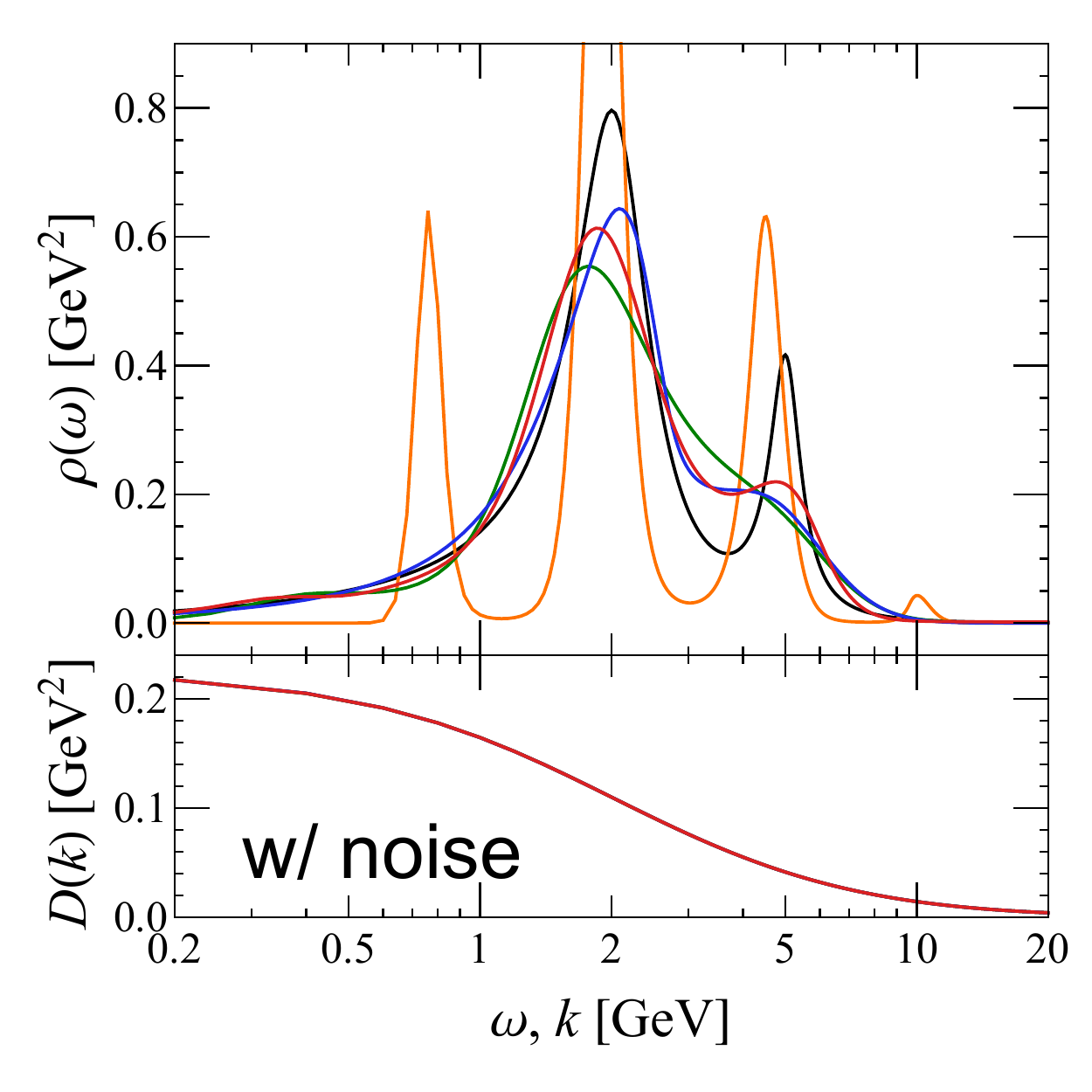}
\includegraphics[width=0.32\textwidth]{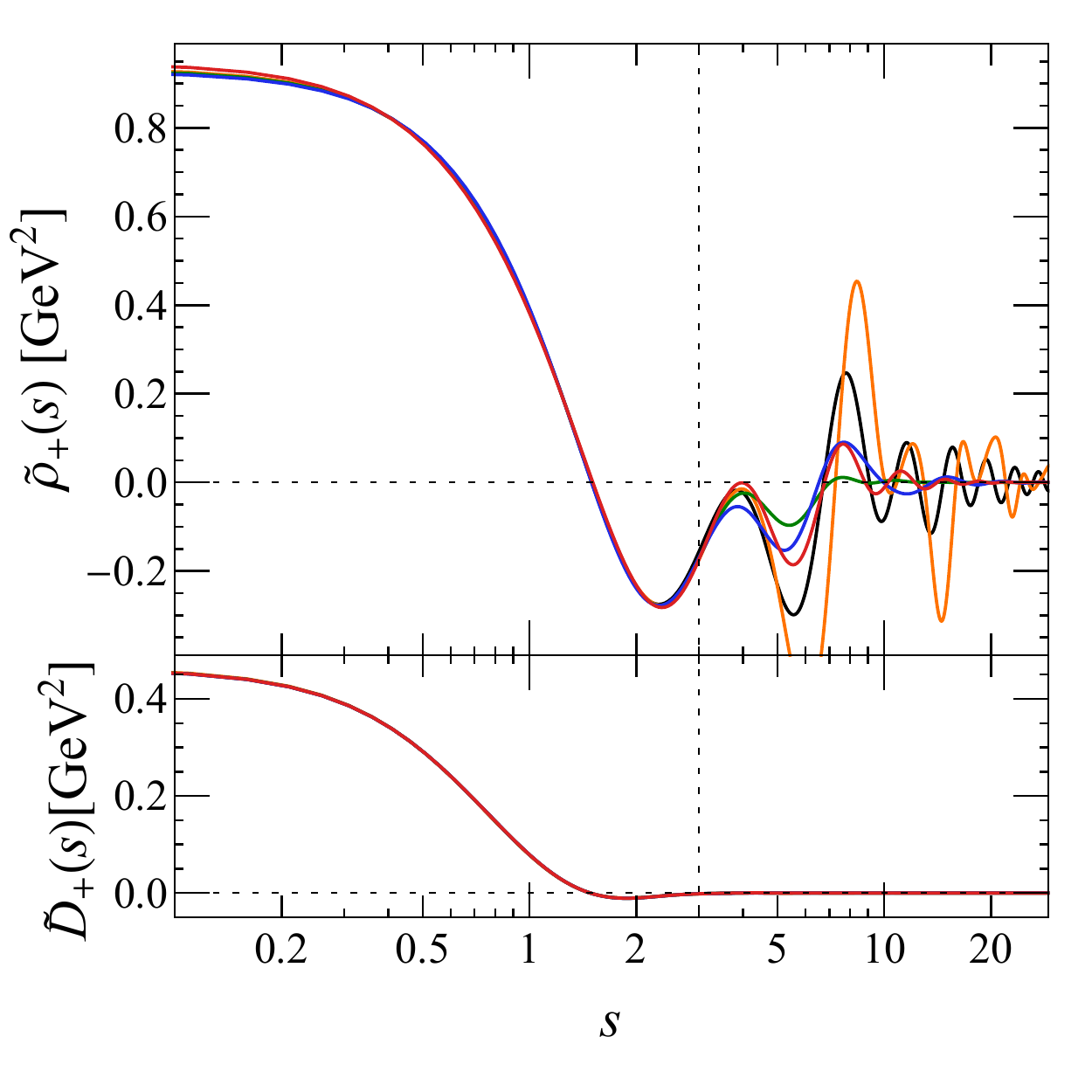}
\includegraphics[width=0.32\textwidth]{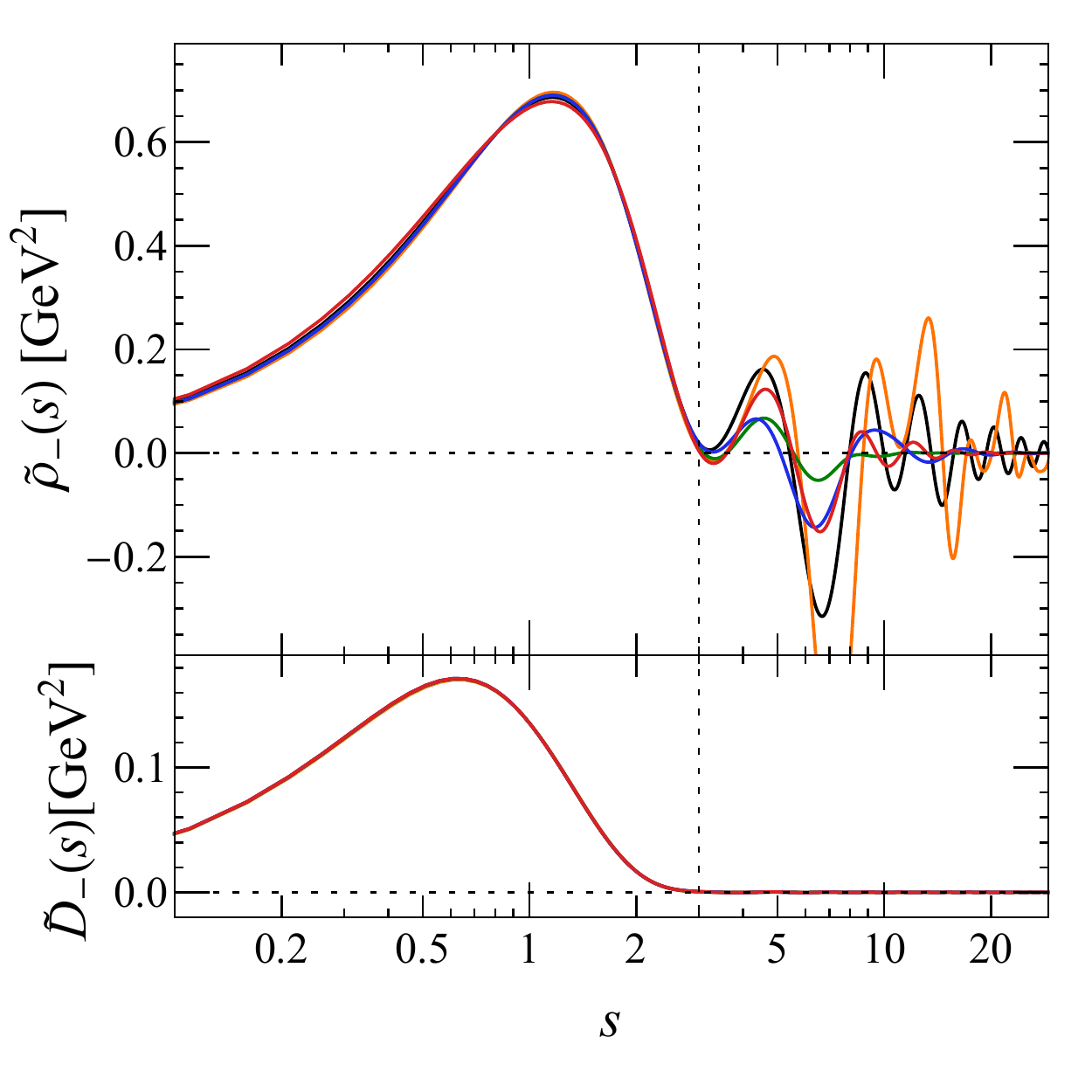}
\caption{Same as Fig.~\ref{fig.performance1} but with noise possessing the variance \eqref{eq.noise}.
\label{fig.performance2}}
\end{figure*}
We start from a known spectral function with two Breit--Wigner peaks
\begin{align}\label{eq.breit_wigner}
    \rho(\omega) = \sum_{n=1}^{2} \frac{4 A_n \Gamma_n \omega}{\left(M_n^{2}+\Gamma_n^{2}-\omega^{2}\right)^{2}+4 \Gamma_n^{2} \omega^{2}}.
\end{align}
with $A_1= 0.8$, $A_2 = 1.0$, $\Gamma_1 = \Gamma_2 = 0.5$~GeV, $M_1= 2.0$~GeV, $M_2 =5.0$~GeV, and compute the corresponding KL correlation functions $D(k_i)$ at $k_i = i\times\Delta k$, with $i=1,2,\cdots,100$, and  $\Delta k=0.2$~GeV.
To investigate the effects of noise in realistic observable data, we prepare two sets of mock data, i.e., with or without random noise on the correlation function, i.e., $\mathrm{D}_i^\text{ideal} = D(k_i)$ and  $\mathrm{D}_i^\text{noisy} = D(k_i) + n_{i}$.
In the latter, we follow Ref.~\cite{Asakawa:2000tr} to set the noise term follow normal distribution $P(n_{i}) = \mathcal{N}(0, \text{var})$ with variance
\begin{align}
    \text{var}_{i} = \big(10^{-4} \times i \times D(k_i)\big)^2\,,
    \label{eq.noise}
\end{align}
Accordingly, the inverse covariance matrix in the $\chi^2$ function~\eqref{eq.chisquare} is chosen to be $C^{-1}_{ij} = \delta_{ij} / \text{var}_i$, for both ideal and noisy data.
Then, we reconstruct the spectral function $\rho(\omega)$ for points $\omega_a = a\times \Delta\omega$, with $a=1, 2, \cdots, 500$, and $\Delta\omega=0.04$~GeV. 

In this subsection, we show the reconstructed spectral functions --- respectively using \texttt{NN} and \texttt{NN-P2P} presentations as well as Maximum Entropy Method --- and their corresponding correlation functions in both generalized coordinate space and generalized momentum space. To investigate the stability property of MEM results against the noise in correlation function and its dependence on the basis truncation, we select two cases: $N_\text{basis}=8$ and $N_\text{basis}=N_k=100$. For \texttt{NN}-architecture, we used 3 hidden layers with width = 64 for each layer. As for the input layer, it is a single constant node set as 1. The output layer contains $N_\omega$ nodes; for \texttt{NN-P2P}-architecture, we used the same set-up for the input layer and hidden layers, but the output layer only has one single node. Besides, all activation functions before output are chosen as \texttt{ELU}. Results are shown in Fig.~\ref{fig.performance1} and ~\ref{fig.performance2}.
We observe that while $\rho(\omega)$'s could be very different in generalized-coordinate space, $\widetilde{\rho}_\pm(s)$'s are alike for generalized-momentum $s\lesssim3$ --- indicated by the vertical dashed lines --- where $\lambda_s$ becomes less than error. This reveals the fundamental difficulty of reconstruction --- with different reconstruction methods, one can always precisely recover the ``low-frequency'' mode of $\widetilde{\rho}$, but the ``high-frequency'' modes are always polluted by the noise/numerical inaccuracy of correlation functions and can never be achieved. Meanwhile, all $D$'s ($\widetilde{D}$'s) are alike, which is automatically guaranteed by the obtained $\chi^2$ function. We have also checked that the relation $\widetilde{D}_\pm(s) = \lambda_s \widetilde{\rho}_\pm(s)$ (cf. Eq.~\eqref{eq.algebra_D_rho}) is fulfilled in the numerical reconstructions. We also note that results using MEM with $N_\text{base}\geq 10$ are consistent with those with $N_\text{base}=100$, for reconstruction of both $\mathrm{D}_i^\text{ideal}$ and $\mathrm{D}_i^\text{noisy}$.
In the comparison of orange curves (MEM with $N_\text{basis}=N_k=100$) in the top and bottom row, it is clear that the MEM result using $N_\text{base}= 100$ is unstable against the noise~\eqref{eq.noise} with the $\{\omega_a\}$ and $\{k_i\}$ set-up in used. The reason is explained in~\ref{sec.mem_unstable}. Note that the instability of MEM with large number of basis does not imply the same issue in other Bayesian methods.

\section{Conclusion}
In this paper, we first analytically solve the eigenvalue problem for continuous K\"allen--Lehmann(KL) convolution and obtain the corresponding eigenfunctions. One can transform both the spectral function and the KL correlations into the eigenfunction space, as a generalized Fourier Transformation. The convolution equation in the generalized coordinate space becomes an algebra equation in the generalized momentum space, and one can formally find the inversion. However, for realistic measurements at which the correlation functions are noisy, the eigenvalue needs to be greater than the noise level so that the corresponding eigenfunction components can be inferred from the data. We find that the magnitude of the eigenvalues can be arbitrarily small, hence the inversion is fundamentally ill-posed. 

We further discuss existing methods to resolve the ill-posedness by introducing regulators. Particularly, the uniqueness of solutions regulated by Artificial Neural Networks, recently proposed by the authors~\cite{Wang:2021jou}, is discussed. From comparing results obtained by different regularization methods, we find that although the spectral function could be different in the generalized-coordinate space, they are consistent with each other in the low generalized-momentum space, at which the eigenvalues are significantly greater than the noise level.

From both analytical analysis and numerical practice, we note that the low generalized-momentum components can be reliably inferred from the observation data of correlation functions. This might point out a new class of prior, which postulates the extrapolation formula for the high generalized-momentum components and rebuild the corresponding spectral function. We leave this for future investigation.

\section*{Acknowledgment} 
We thank Drs. Heng-Tong Ding, Swagato Mukherjee and Gergely Endr\"odi for helpful discussions. The work is supported by (i) the BMBF under the ErUM-Data project (K. Z.), (ii) the AI grant of SAMSON AG, Frankfurt (K. Z. and L. W.), (iii) Xidian-FIAS International Joint Research Center (L. W), (iv) Natural Sciences and Engineering Research Council of Canada (S. S.), (v) the Bourses d'excellence pour \'etudiants \'etrangers (PBEEE) from Le Fonds de Recherche du Qu\'ebec - Nature et technologies (FRQNT) (S. S.), (vi) U.S. Department of Energy, Office of Science, Office of Nuclear Physics, grant No. DE-FG88ER40388 (S. S.). K. Z. also thanks the donation of NVIDIA GPUs from NVIDIA Corporation.

\begin{appendix}
\section{Regularized Inverse K\"allen--Lehmann Kernel}
\label{sec.reg_BackusGilbert}
In this section, we discuss the characteristics of the integral~\eqref{eq.inverse_kl_integral} involved in the inverse K\"akken--Lehmann transformation,
\begin{align}
I(x) \equiv    \int_{-\infty}^{\infty} \cos(s x)\cosh(\pi s/2)  \,\mathrm{d}s\,,
\end{align}
Obviously, such an integral does not converge, and we define the regularized integral as
\begin{align}
\begin{split}
I_{\varepsilon}(x) 
\equiv\;&   
    \int_{-\infty}^{\infty} \cos(s x)\cosh(\pi s/2) e^{-\frac{\varepsilon^2 s^2}{2}} \,\mathrm{d}s
=
    2\pi e^{\frac{\pi^2}{8\varepsilon^2}}
    \cos\Big(\frac{\pi x}{8\varepsilon^2}\Big)
    \frac{e^{-\frac{x^2}{2\varepsilon^2}}}{\sqrt{2\pi}\varepsilon}\,,
\end{split}
\end{align}
We note that
\begin{align}
    \int_{-\infty}^{\infty}
    (x-x_0)^{2n+1}
    I_{\varepsilon}(x-x_0) \mathrm{d}x
=\,&
    0\,,
\\
    \lim_{\varepsilon\to0} \int_{-\infty}^{\infty}
    (x-x_0)^{2n}
    I_{\varepsilon}(x-x_0) \mathrm{d}x
=\,&
    2\pi\,\Big(-\frac{\pi^2}{4}\Big)^{n}\,,
\end{align}
Hence,
\begin{align}
\begin{split}
&    \lim_{\varepsilon\to0}
    \int_{-\infty}^{\infty}
    f(x)
    I_{\varepsilon}(x-x_0) \mathrm{d}x
\\=\,&
    \lim_{\varepsilon\to0}
    \int_{-\infty}^{\infty}
    \sum_{n=0}^{\infty}
    \frac{f^{(n)}(x_0)}{n!}(x-x_0)^n
    I_{\varepsilon}(x-x_0) \mathrm{d}x
\\=\,&
    \lim_{\varepsilon\to0}
    \int_{-\infty}^{\infty}
    \sum_{n=0}^{\infty}
    \frac{f^{(2n)}(x_0)}{(2n)!}(x-x_0)^{2n}
    I_{\varepsilon}(x-x_0) \mathrm{d}x
\\=\,&
    2\pi\,\sum_{n=0}^{\infty}
    \frac{(-\pi^2/4)^n}{(2n)!}f^{(2n)}(x_0)\,.
\end{split}
\end{align}

Assuming that the order of $\int dx$ and $\lim_{\varepsilon\to0}$ are interchangeable, 
we reach that
\begin{align}
\begin{split}
&   \int_{-\infty}^{\infty}
    f(x)
    I(x-x_0) \mathrm{d}x
=
    2\pi\,\sum_{n=0}^{\infty}
    \frac{(-\pi^2/4)^n}{(2n)!}f^{(2n)}(x_0)\,.
\end{split}
\end{align}
If we know the analytic continuation of $f(x)$ in the complex plan, we can further simplify the result as
\begin{align}
\begin{split}
&   \int_{-\infty}^{\infty}
    f(x)
    I(x-x_0) \mathrm{d}x
=
    \pi\,\Big(f(x_0+\frac{\pi i}{2})
    + f(x_0-\frac{\pi i}{2}) \Big)\,.
\end{split}
\end{align}

Exploiting these properties, we can re-derive the well-known optical theorem~\cite{zee:2010quantum}
\begin{align}
\begin{split}
\rho(\omega) 
=\,&
    \int_0^{\infty} \frac{D(k) \, \mathrm{d}k}{\pi \, \omega}
    I\Big(\ln\frac{k}{a} - \ln\frac{\omega}{a} \Big)
\\=\,&
    \int_{-\infty}^{\infty} \frac{D(a\,e^x) a\,e^x \, \mathrm{d}x}{\pi \, \omega}
    I\Big(x - \ln\frac{\omega}{a} \Big)
\\=\,&
    -2\,\text{Im}[D(i\omega)]\,,
\end{split}
\label{eq.optical_de}
\end{align}
and the zero-frequency limit which relates to the transport coefficients.
If we start from the correlation function defined in the complex plane~\eqref{eq.optical_de}, we find
\begin{align}
\begin{split}
&    \lim_{\omega \to 0} \frac{\rho(\omega)}{\omega}
\\=\,&
    - \lim_{\omega \to 0} \frac{D(i\omega)-D(-i\omega)}{i \omega}
\\=\,&
    -2\lim_{\omega \to 0} \sum_{n=0}^{\infty} 
    \frac{ (-1)^n \omega^{2n}}{(2n+1)!} D^{(2n+1)}(\omega)
\\=\,&
    -2 D'(0)\,.
\end{split}
\end{align}
It shall be worth noting that such a relation can be derived even if restricting $f$ to be defined on the real axis. Noting that
\begin{align}
\begin{split}
&    \frac{\partial^{2n}[D(a\,e^x) a\,e^x]}{\partial x^{2n}}
\bigg|_{x=\ln\frac{\omega}{a}} 
\\=\,&
    \Big(\omega \frac{\partial}{\partial \omega} \Big)^{2n} \Big[\omega D(\omega)\Big]
\\=\,&
    D(\omega)\omega + (2^{2n}-1) D'(\omega) \omega^2 + \mathcal{O}(\omega^3)\,,
\end{split}
\end{align}
and
\begin{align}
&\sum_{n=0}^{\infty} \frac{1}{(2n)!} 
\bigg(\!\!-\frac{\pi^2}{4}\bigg)^n
= 0\,,\\
&\sum_{n=0}^{\infty} \frac{2^{2n}-1}{(2n)!} \bigg(\!\!-\frac{\pi^2}{4}\bigg)^n
= -1\,,
\end{align}
we can start from the real-axis relation~\eqref{eq.delta_drv} and reach that
\begin{align}
\begin{split}
&    \lim_{\omega \to 0} \frac{\rho(\omega)}{\omega}
=
    \lim_{\omega \to 0} 
     \frac{2}{\omega^2}\,\sum_{n=0}^{\infty}
    \frac{(-\pi^2/4)^n}{(2n)!}
    \frac{\partial^{2n}[D(a\,e^x) a\,e^x]}{\partial x^{2n}}\bigg|_{x=\ln\frac{\omega}{a}}
=
    -2 D'(0)\,.
\end{split}
\end{align}

\section{Stability Properties of MEM Results Against the Noise in Correlation Functions}
\label{sec.mem_unstable}
\begin{figure}[!h]
    \centering
    \includegraphics[width=0.5\textwidth]{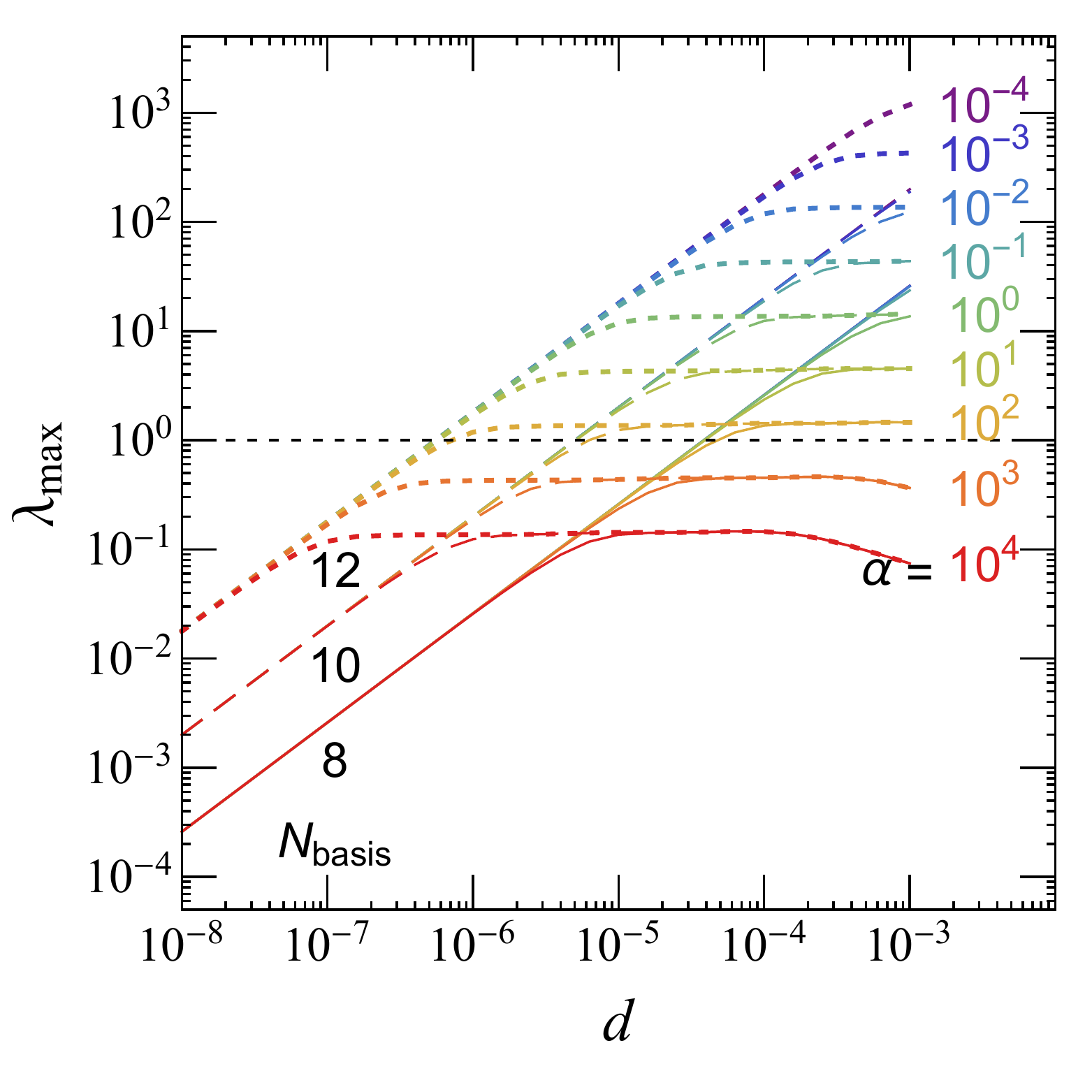}
    \caption{Greatest eigenvalue of $M^{-1}$ as functions of noise level $d$. Top(purple) to bottom(red) represent results for $\alpha = 10^{-4}$ to $10^{4}$. Solid, dashed, and dotted curves correspond to truncation of basis $N_\text{basis}=8$, $10$, and $12$, respectively.}
    \label{fig.eigenvalue}
\end{figure}
In this Appendix, we discuss the stability of MEM solution~\eqref{eq.coef_mem} against the noise of correlation functions.
Given $\alpha$, the solution of MEM satisfies the self-consistent equation:
\begin{align}
\rho_a =\,& 
    \text{DM}_a \exp\Big(\sum_{b} c_b U_{ab} \Big)\,,\\
\alpha\, c_a 
   =\,& 
   \sigma_a \sum_{i}
   \frac{D^\text{obs}_i-\sum_{c,d}\sigma_d U_{cd}V_{id}\rho_c\Delta\omega}{d_i^2} V_{ia}\,,
\end{align}
where the inverse covariance matrix in the $\chi^2$ function~\eqref{eq.chisquare} has been chosen to be $C^{-1}_{ij} = \delta_{ij} d_i^{-2}$.
To study the stability of MEM, we solve the change in coefficients ($\delta c$) against the change in correlation function ($\delta D^\text{obs}$), which satisfies the equation
\begin{align}
\begin{split}
&    \alpha\, \delta c_a 
   + \sigma_{a}\sum_{i,c,b,d}\sigma_{d}
    \frac{V_{ia}V_{id}}{d_i^2} \rho_c U_{cb} U_{cd} \delta c_b \Delta\omega
=
    \sigma_{a} \sum_i 
    \frac{\delta D^\text{obs}_i}{d_i^2} V_{ia}\,.
\end{split}
\end{align}
Assuming uniform error\footnote{We have checked that stability property are qualitatively the same for both uniform and non-uniform errors.}, $d_i=d$, we can further simplify the equation as
\begin{align}
   \sum_b M_{ab} \,\delta c_b
   = \sum_i 
   \frac{\delta D^\text{obs}_i}{d} V_{ia}
   \,,
\end{align}
hence
\begin{align}
   \delta c_b
   = \sum_{i,a} (M^{-1})_{ba}
   \frac{\delta D^\text{obs}_i}{d} V_{ia}
   \,,
\end{align}
where $a,b=1,2,\cdots,N_\text{basis}$, with $N_\text{basis}$ being the number of included basis, and
\begin{align}
M_{ab} \equiv\;&\frac{\alpha\,d}{\sigma_a}\, \delta_{ab} 
   + \Delta\omega \frac{\sigma_a}{d} \sum_{c}U_{ca} U_{cb} \rho_c\,.
   \label{eq.mat}
\end{align}
Noting that $V_{ia}\sim\mathcal{O}(1)$, $\delta D^\text{obs}_i \sim \mathcal{O}(d)$, henceforth $\frac{\delta D^\text{obs}_i}{d} \sim \mathcal{O}(1)$, the order of magnitude of $\delta c$ shall be the same as that of the greatest eigenvalue of $M^{-1}$. 
We use the set-up in Sec.~\ref{sec.regulator.practical}, and substitute $\rho$-vector in Eq.~\eqref{eq.mat} by the ground truth therein~\eqref{eq.breit_wigner}, and compute numerically the greatest eigenvalue of $M^{-1}$, $\lambda_\text{max}$. 
When $\lambda_\text{max} \gg 1$, the difference in correlation would be strongly amplified and therefore become unstable. 
Roughly speaking, the solution's stability criterion is that whether $\lambda_\text{max} \lesssim 1$ is satisfied or not.
In Fig.~\ref{fig.eigenvalue}, we show $\lambda_\text{max}$ at varies noise ($d$), regulator factor ($\alpha$), and number of basis ($N_\text{basis}$). As expected, $\lambda_\text{max}$ decreases when noise decreases or $\alpha$ increases, i.e., putting more weights on the entropy term. For noise level $d\sim10^{-4}$ --- correspond to Eq.~\eqref{eq.noise} --- and regulator parameter $\alpha\sim 10^{-2}$, --- at which that $P(\alpha|D,\text{DM})$ takes its maximum --- we observe that $\lambda_\text{max}\sim 1$ when $N_\text{basis}=8$ whereas $\lambda_\text{max}\sim 20$ when $N_\text{basis}=10$, which explains the stability(instability) of the former(latter). We find qualitatively the same behavior for $\lambda_\text{max}$ when taking the $\rho_c$ in Eq.~\eqref{eq.mat} to be the one extracted from noiseless data or the spectral functions with different values for masses/widths or number of peaks.

\end{appendix}

\bibliographystyle{elsarticle-num}
\bibliography{ref}
\end{document}